# Impact of the coagulation of dust particles on Mars during the 2018 global dust storm

*T. Bertrand*[*a,b], *M. Kahre*[b], *R. Urata*[c], *A. Määttänen*[d], *F. Montmessin*[e], *J. Wilson*[b], *M. Wolff*[f]
* corresponding author
e-mail: tanguy.bertrand@obspm.fr

**Affiliations:**
[a] Laboratoire d'Etudes Spatiales et d'Instrumentation en Astrophysique (LESIA), Observatoire de Paris, Université PSL, CNRS, Sorbonne Université, Univ. Paris Diderot, Sorbonne Paris Cité, 5 place Jules Janssen, 92195 Meudon, France
[b] Ames Research Center, Space Science Division, National Aeronautics and Space Administration (NASA), Moffett Field, CA, USA
[c] Bay Area Environmental Research Institute, Moffett Field, CA
[d] LATMOS/IPSL, Sorbonne Université, UVSQ Université Paris-Saclay, CNRS, Paris, France
[e] LATMOS/IPSL, UVSQ Université Paris-Saclay, Sorbonne Université, CNRS, Guyancourt, France
[f] Space Science Institute, Boulder, CO USA

- We explored the impact of dust coagulation in a Mars Global Climate Model during a global dust storm.

- Dust coagulation significantly increases the atmospheric effective dust particle size during the storm

- Due to coagulation, the upper atmosphere tends to be more depleted in dust and up to 20 K colder during the mature and decay phases of the storm

- In general, the representation of the decay phase of the storm relative to MCS dust observations is improved

- Coagulation is an important process to consider when simulating dust storms on Mars or Mars at high obliquity

# Abstract

Coagulation of particles occurs when two particles collide and stick together. In the Martian atmosphere, coagulation of dust would increase the dust effective particle size, as small particles accrete to larger particles. Murphy et al. (1990) concluded that Brownian coagulation of dust in the Martian atmosphere was not significant, due to the low dust particle mixing ratios, while Montmessin et al. (2002) and Fedorova et al. (2014) showed that it mostly involves dust particle radii smaller than 0.1 µm. However, the effects of coagulation have never been explored in 3D, during a global dust storm, and in presence of larger numbers of small particles. Here we revisit this issue by using the NASA Ames Mars Global Climate Model (MGCM) to investigate the temporal and spatial changes in dust particle sizes during the 2018 global dust storm due to dust coagulation and the overall impact of these processes on Mars' climate. Our parameterization for dust coagulation includes the effect of Brownian motion, Brownian diffusion enhancement, and gravitational collection. We show that Brownian motion and Brownian diffusion enhancement dominate gravitational collection. Coagulation has a significant impact during the global storm, with coagulation rates increased by a factor of 10 compared to non-storm conditions. The mean effective particle radius can be increased by a factor of up to 2 due to coagulation, leading to a 20 K colder atmosphere above 30 km altitude. Overall, our parameterization improves the representation of the decay phase of the storm relative to MCS dust observations. Coagulation also remains a significant process affecting dust outside the storm period if large numbers of submicron-sized particles are involved. As coagulation removes the small sub-micron particles within a relatively short time, it may therefore be possible, in GCMs, to lift larger amounts of submicron-sized particles from the surface without excess dust buildup in the atmosphere, thus improving the agreement with some of the observations without diverging from the observed column opacities.

# 1. Introduction

Regional and global dust storms (GDSs) on present-day Mars dominate weather and climate variability, including diurnal pressure variations, atmospheric heating, and global circulation (e.g. Zurek and Martin, 1993; Newman et al., 2002; Strausberg et al., 2005). This is due to the strong radiative impact of dust particles (absorption and scattering of visible and infrared radiation) on the thermal state of the thin Martian atmosphere (e.g., Gierasch and Goody, 1972, Haberle, 1986, Read et al., 2015, Wolff et al., 2017) and the role of dust as condensation nuclei for water and $CO_2$ ice cloud particles, which also affect radiative fluxes (e.g., Navarro et al., 2014, Kahre et al., 2015).

GDSs are the most dramatic of these dust events since they enshroud Mars in a veil of dust for months, producing dust optical depths $\tau > 3$ in the visible bands, or $\tau > 0.5$ in the infrared (IR) at 9.3 µm over large areas of the planet (which corresponds to a tenfold increase compared to non-storm conditions; Cantor, 2007; Montabone et al., 2015, 2020, Lemmon et al., 2019). They typically occur about three times per Mars decade, but with great irregularity. To date, they have all been observed to occur in the second half of the Mars year (solar longitude Ls=180-360°), with large variability in onset location and timing.

Studying these storms has major implications for understanding the present climate, modeling the past climates (including past water cycles and potential habitability), interpreting the geology, and quantifying and mitigating the risks storms pose to Mars missions. The most recent of these events, the 2018 GDS (Mars Year 34), was the most comprehensively observed GDS in history with two rovers on the surface and six spacecraft in orbit, in conjunction with terrestrial observations.

The analysis of the different datasets and their interpretation with atmospheric models have allowed for a better characterization and understanding of GDSs, (Guzewich et al., 2019, Smith, 2019, Montabone et al., 2020, Wolkenberg et al., 2020), how they expand through dust lifting and transport (Gillespie et al., 2020, Bertrand et al., 2020), how they impact the water cycle (Aoki et al., 2019, Heavens et al., 2019, Rossi et al., 2021) and more generally the lower atmosphere (Kleinboehl et al. 2020, Bertrand et al., 2020) and the upper atmosphere (e.g. Girazian et al., 2020, and other publications in the same special issue).

Bertrand et al. (2020) performed a modeling study of the 2018 GDS, using the NASA Ames Mars Global Climate Model (MGCM), and made a detailed characterization of the storm's formation, evolution and decay, as well as of the atmospheric temperatures, winds, and dust and cloud abundances. This modeling effort used a spatially uniform and temporally constant size distribution for lifted dust. In the atmosphere, the simulated mean effective dust particle size $r_{eff,atm}$ increased during the storm, due to the strong convective movements and vertical mixing transporting large particles at high altitude. However, it was difficult to reproduce the rapid onset of the storm and the strong decay rate in opacity, which suggested that another mechanism increasing even more the dust particle size throughout the storm's evolution was not taken into account or not well reproduced by the model (See Section 2, Background on dust particle sizes during Martian dust events). This inconsistency led us to explore options to make modeled particles even bigger and thus to envision coagulation of dust, which occurs when two particles collide and stick together, as a plausible mechanism. This mechanism would shift the atmospheric dust particle distribution towards a larger effective radius during the GDS, as the small particles accrete to larger particles.

We note that other mechanisms could take place during storms and also produce a general increase in particle sizes, such as, for instance, 1) changes in lifted particle size at the surface due to different active reservoirs, 2) depletion of small particles as the storm increases in intensity, or 3) increase in surface wind stress allowing for larger particles to be lifted. However, in this paper we only focus on coagulation and we leave the investigation of these processes to a future work.

Previous studies showed that Brownian coagulation on Mars quickly removes particles with $r < 0.1$ µm and number density $> 1000$ cm$^{-3}$ (Michelangeli et al. 1993, Montmessin et al. 2002, Fedorova et al., 2014; see Section 3.2). However, the efficiency of the different coagulation types (e.g., Brownian vs. gravitational coagulation) has never been investigated with a 3D GCM in the context of a GDS, when atmospheric dust abundance (i.e., dust particle number densities) is significantly enhanced. This paper aims to fill this knowledge gap.

Here we focus on dust coagulation during the 2018 GDS, by implementing the equations of coagulation of Jacobson (2005) in the MGCM and simulating the dust storm with and without active dust coagulation. Our science questions include: What is the impact of dust coagulation on dust particle size and vertical distribution and on the column opacities during a major event such as the 2018 GDS? What are the coagulation rates? Does coagulation affect the atmospheric state (temperatures, winds)?

In Section 2, we provide a background about previous works related to dust coagulation on Mars and previous observations of dust particle sizes during Martian dust storms. In Section 3, we present the equations used to parameterize different types of dust coagulation in the MGCM, and we compare characteristic properties of coagulation in Martian conditions for different dust particle sizes. Section 4 presents our modeling results. It includes a short description of the Ames Mars global climate model and of the implementation of the coagulation equation, numerical tests in 0-D (in a finite volume) on lognormal particle size distributions, and the main 3D simulations of the 2018 global dust storm. Sections 5 and Section 6 include a general discussion and a summary of the results, respectively.

# 2. Background on dust particle sizes during Martian dust events

The parameters of the Martian atmospheric dust particle size distributions (effective particle size $r_{eff}$ and variance $v_{eff}$) strongly impact the radiative transfer in Mars' atmosphere and are therefore important drivers of Mars' climate (Hansen and Travis, 1974). They can be retrieved directly from observations of the atmosphere (specifically the wavelength-dependent way it scatters, absorbs, and/or emits radiation). Typical values during non-storm events have been reported, yielding an effective radius (of equivalent volume sphere) of ~1-1.5 µm (+/- 50%) and variances around 0.2–0.5 from a broad set of spacecraft and instruments at the surface and in orbit, e.g., Mariner 9, Viking, Pathfinder (Pang et al., 1976; Toon et al., 1977; Clancy & Lee, 1991, Pollack et al., 1995; Ockert-Bell et al., 1997; Tomasko et al., 1999), Mars Exploration Rovers (MER; Lemmon et al., 2004), Mars Global Surveyor Thermal Emission Spectrometer (Clancy et al., 2003; Wolff and Clancy, 2003; Wolff et al., 2006; Clancy et al., 2010), Mars Express (Määttänen et al., 2013; Rannou et al., 2006), Mars Reconnaissance Orbiter (Guzewich et al., 2014; Wolff et al., 2009), and Mars Science Laboratory (MSL; Chen-Chen et al., 2019; McConnochie et al., 2018; Vicente- Retortillo et al., 2017, Lemmon et al., 2019). For more details, a summary of observationally derived $r_{eff}$ and $v_{eff}$ can be found in Dlugach et al. (2003), with more recent efforts included in Smith (2008) and Kahre et al. (2017, Book Chapter).

An increase in dust particle size during major dust events is a general trend supported by several observations and studies (e.g., Kahre et al., 2008, Elteto and Toon, 2010, Smith and Wolff, 2014; Vicente-Retortillo et al., 2017). For instance, a correlation between dust opacity and particle sizes was shown by Smith and Wolff (2014) and Lemmon et al. (2015), using MER Navcam and Pancam datasets, respectively. At the peak of the 2001 GDS, TES-derived particle effective radii ranged from approximately 1.0 to 2.5 µm (Wolff and Clancy, 2003; Clancy et al., 2003). More recently, Lemmon et al. (2019) and Chen-Chen et al. (2021) used several instrument and navigation camera datasets from MSL to retrieve atmospheric dust particle sizes at Gale crater before and during the 2018 GDS (Mars Year 34). They showed a similar positive correlation between dust opacity and particle sizes, with an abrupt increase from about 1.2-1.5 to >4 µm in effective radius during the onset and expansion phase of the storm, and values remaining >3 µm for most of the decay phase before declining to seasonal values.

Note that these values represent an average over properties in the line of sight and assume a specific vertical dust distribution that is generally not well constrained by observations. In addition, several studies show that the dust size distribution is not well represented by a single particle size but is better fit with bi-modal distributions (e.g., Fedorova et al., 2014). In addition to the particle sizes, the different particle compositions and shapes also govern their interaction with radiation in the atmosphere which complicates the retrievals (e.g., Lemmon et al., 2015; Smith, 2008; Wolff et al., 2006).

Climate models typically assume a surface dust reservoir with a fixed distribution of particle sizes (i.e. with a fixed effective particle radius), with size evolution occurring in the atmosphere because of horizontal and vertical transport, gravitational sedimentation, and interactions with water ice clouds. They typically use a lifted dust effective radius $r_{eff,lift}$ of 2-3 µm in order to simulate the atmospheric dust distribution with a column mean effective radius $r_{eff,atm}$ of ~1.5 µm (e.g., Kahre et al., 2008, Wang et al., 2018, also see review in Haberle et al., 2019).

Nevertheless, an increase in the effective dust particle size $r_{eff,atm}$ can be expected during dusty conditions, when the timescales for the convective movements and the vertical mixing overcome those of gravitational sedimentation (e.g., Kahre et al., 2008). Typically, the downward motion due to gravity of a 1 µm particle at 20 km altitude can be balanced by an updraft of ~1 cm s$^{-1}$. Such vertical wind speeds (and even higher values) can be caused by many different processes. Two examples are dust storms (and notably "rocket dust storms", Spiga et al., 2013) and upward transport of the rising branch of the Hadley circulation (Haberle et al., 1982; Murphy et al., 1990, 1993; Kahre et al., 2008), but many others exist. Strong updrafts associated with plumes of dust during the 2018 GDS have also been modeled in the MGCM (Bertrand et al., 2020) and confirmed by observations ("dust towers", Heavens et al., 2019).

The decay phase of global dust storms is particularly well suited to analyze the processes related to the evolution of dust particle sizes since lifting of dust is limited during this phase (e.g., Fig. 9 in Bertrand et al., 2020). For instance, the particle size distribution inferred by Toon et al. (1977) from Mariner 9 data remained essentially unchanged at southern subtropical latitudes during the decay of the 1971-1972 GDS, suggesting that a process other than gravitational sedimentation was at work to maintain a constant dust particle effective radius. The particle sizes retrieved from MSL instruments during the 2018 GDS (Lemmon et al., 2019, Chen-Chen et al., 2021) also showed little change during most of the decay phase of this storm.

# 3. Coagulation: background, theory, equations and key parameters

In this section, we present the coagulation equations implemented in the MGCM, and compare the theoretical atmospheric coagulation rates and timescales for different coagulation types in Martian conditions. The reader is referred to Jacobson (2005) and Seinfeld and Pandis (2006) for more details.

## 3.1. Definitions

Coagulation refers to the production of particles whose size results from the collision and subsequent sticking together of smaller particles. This process is thus a sink for the smallest particles of a size distribution and a source for the larger particles. It conserves the volume concentration and the mass in the atmosphere, while the number concentration is reduced. Different types of coagulation exist depending on the type of collision involved:

(1) **Brownian coagulation** involves collisions between particles resulting from their Brownian motion, i.e., their random motion in the background gas due to their irregular bombardment by gas molecules. It is sometimes referred to as thermal coagulation.
(2) **Convective Brownian diffusion enhancement** (also referred throughout the paper as Brownian diffusion-enhanced coagulation) occurs when a large particle falls through the atmosphere, and creates eddies and turbulences in its wake, thus increasing the diffusion of other particles to its surface and further enhancing the particle coagulation.
(3) **Gravitational coagulation** involves vertical (downward) collisions caused by the different sedimentation velocities of particles of different size (the largest one catches up to the smallest one).

The coagulation can also be enhanced by **turbulent inertial motion** (velocity gradients in turbulent flows cause relative particle motion, i.e. acceleration is a function of particle size) and **turbulent shear** (wind shear in turbulent air would enhance collisional rates). However, as shown in Section 3.4 these two coagulation types are negligible on Mars, and throughout this paper we will therefore focus on the impact of Brownian, Brownian diffusion-enhanced, and gravitational coagulation on dust particles on Mars.

## 3.2. Previous studies of dust coagulation on Mars

Many studies to date have described numerical techniques for solving the differential coagulation equation (Rossow, 1978, Jacobson 2002, 2005, and references therein). As shown in Section 3.3, the coagulation process strongly depends on the particle number densities and radii involved. Rossow (1978) calculated the timescale for Brownian coagulation of dust particles on Mars to be ~$10^7$ s (3.8 Earth months), i.e., about the lifetime of a major Martian dust storm. Based on the timescales for sedimentation and mixing of dust particles, he concluded that Brownian coagulation during a storm could produce micron-sized dust particles (at the expense of submicron particles) at a rapid enough rate to offset their preferred sedimentation, thus allowing for a constant effective dust particle size during the decay phase of a dust storm (as suggested by the observations of the 1971 and 1977 global dust storms).

Murphy et al. (1990) investigated dust coagulation further by using the 1D and 2D aerosol models of Toon et al. (1988) and a modified-gamma dust particle size distribution. The model includes several physical mechanisms that influence upon the suspended dust, such as sedimentation, diffusion, surface deposition, resolution of an input particle size distribution as well as Brownian coagulation, based on the equations of Rossow (1978). By simulating the decay of a Martian global dust storms with and without the coagulation active, the authors concluded that Brownian coagulation has only a slight effect upon the decline of the optical depth during the decay phase of the storm, with a decrease in opacity of only 4% after 110 sols with active coagulation compared to the results without coagulation.

Friedlander (1977), Montmessin (2002), Montmessin et al. (2002) and Fedorova et al. (2014) showed that the coagulation of dust particles on Mars involves particles radii smaller than 0.1 μm, leaving larger particles essentially unaffected. In particular, they showed that the coagulation of these small particles after several days results in particle size distributions that are approximately lognormal. They found that modified-gamma dust particle size distributions, involving large amount of small particles and suggested on Mars by several studies (e.g. Toon et al., 1977, Clancy et al., 1995), are highly unstable and should quickly (after a few sols) and converge towards a lognormal distribution due to the influence of Brownian coagulation removing the small particles (see Fig. 2 in Montmessin et al., 2002). Among other reasons, this motivates the use of lognormal function to describe dust distributions in climate models.

Fedorova et al. (2014) explored the effect of Brownian coagulation on an observed bi-modal particle size distribution by solving the coagulation equation based on the semi-implicit scheme described in Jacobson (2005; Jacobson et al., 1994). They found that coagulation of small particles could explain the observed bi-modal type of particle size distribution. In particular, the observed small mode with $r_{eff}$ = 0.04–0.05 μm and a number density N>1000 $cm^{-3}$ is found to be highly unstable (at a timescale of less than a week) due to coagulation and requires a source of small particles to balance the loss by coagulation. Another observed small mode with $r_{eff}$ = 0.07 μm and N < 100 $cm^{-3}$ is found to be more stable and can survive 50-100 sols or more.

## 3.3. Equations of coagulation

### 3.3.1. Binned particle size distribution

The coagulation equation can be discretized in terms of a monomer size distribution, with bins of particles for each monomer size. In this distribution, each volume bin is a multiple of the smallest volume. Thus, coagulation always results in a particle moving to a specific bin instead of being partitioned between two bins. This volume-ratio size discretization specifies that the volume of each particle in size bin k equals the volume of a particle in the smallest bin multiplied by k, and can be written as:

$$V_k = kV_1$$

$$r_k = V_{rat}^{1/3} r_{k-1} \quad \text{with} \quad V_{rat} = \left(\frac{r_{N_B}}{r_1}\right)^{\frac{3}{N_B-1}} \tag{1}$$

With $V_k$ and $r_k$ the volume (m³) and radius (m) of one particle of the bin k, $N_B$ the total number of size bins ($V_1$ and $V_{NB}$ are the smallest and largest volume bins respectively), and $V_{rat}$ the constant volume ratio.

### 3.3.2. Semi-implicit scheme for solving coagulation equations

We solve the coagulation equations based on the semi-implicit scheme described in Jacobson (2005) and Jacobson et al. (1994) with conservation of volume (which coagulation physically does) and in terms of number concentration (n, in particles m⁻³):

$$n_{k,t} = \frac{n_{k,t-h} + \frac{h}{V_k}\sum_{j=1}^{k}(\sum_{i=1}^{k-1} f_{i,j,k} V_i \beta_{i,j} n_{i,t} n_{j,t-h})}{1 + h\sum_{j=1}^{N_b}(1 - f_{k,j,k})\beta_{k,j} n_{j,t-h}} \tag{2}$$

With h the time step (s), subscripts t and t-h the final and initial times, subscripts i,j,k the size bin indices, $V_k$ the volume of one particle of size bin k, β the coagulation kernel of the two colliding particles (m³ particle⁻¹ s⁻¹; it is a coefficient related to the coagulation rate that depends on the particle radii, see below) and f the volume fraction (volume of an intermediate particle partitioned between two bins), defined in Jacobson et al. (2005). The reader is referred to this paper and to Appendix A for more details.

The numerator of Equation 2 corresponds to the coagulation production rate of particles, while the denominator corresponds to the loss rate. This semi-implicit scheme conserves mass and volume. By increasing the resolution of the binned size distribution ($N_B$), the error in number concentration approaches zero. In general, we use $N_B$=40 bins, which gives satisfactory results (see Section 4).

### 3.3.3. Coagulation and collision kernels

The coagulation kernel $β_{i,j}$ in Equation 2 is a coefficient related to the coagulation rate between the two colliding particles i and j, and is therefore a key parameter to describe the process of coagulation. It is expressed in m³ particle⁻¹ s⁻¹ and can sometimes be written as a product between a sticking (or coalescence) efficiency coefficient $E_{coal}$ and a collision kernel K:

$$\beta_{i,j} = E_{coal} K_{i,j} \tag{3}$$

Individual kernels for each coagulation type can be summed to compute the main coagulation kernel. In this section, we summarize the equations defined by Jacobson et al. (2005) that describe the collision kernels of different coagulation types applicable to dust particles on Mars, and we compare their magnitude for both Martian and terrestrial conditions.

#### 3.3.3.1. Brownian motion

We use the equation for the coagulation kernel for Brownian motion $K^B$ as defined by Jacobson et al. (2005) for the transition regime, which is appropriate for the case of dust particles on Mars, as shown in Appendix

B. The equation simplifies to that in the continuum regime for small Knudsen numbers (i.e. for large, > 10 μm dust particles on Mars) and to that in the free molecular regime for large Knudsen numbers (i.e., for small, < 0.1 μm dust particles on Mars):

$$K_{i,j}^B = \frac{4\pi(r_i + r_j)(D_{p,i} + D_{p,j})}{\frac{r_i + r_j}{r_i + r_j + (\delta_i^2 + \delta_j^2)^{1/2}} + \frac{4(D_{p,i} + D_{p,j})}{(r_i + r_j)(\varphi_{p,i}^2 + \varphi_{p,j}^2)^{1/2}}} \quad (4)$$

Where $r_i$ and $r_j$ are the radii of particles i and j, respectively, $D_{p,i}$ and $D_{p,j}$ are their Brownian diffusion coefficients (m$^2$ s$^{-1}$, modified with the Cunningham slip-flow correction factor G), $\delta_i$ and $\delta_j$ are the mean distances related to their mean free path, and $\varphi_{p,i}$ and $\varphi_{p,j}$ are their thermal velocities, as defined in Jacobson et al. (2005) and summarized in Appendix A.

Appendix Figure B-1 shows the evolution of the Brownian diffusion coefficient (which represents the Brownian motion) versus the particle size. In Martian conditions, the enhanced flow of small particles leads to a higher Knudsen number and thus a higher coefficient diffusion than that in terrestrial conditions for the same small particles, by a factor ~100. For large particles, the Knudsen number is reduced, and the diffusion coefficient is of the same order of magnitude in both conditions (it slightly differs due to the difference between Martian and terrestrial atmospheric temperature T and dynamic viscosity η).

### 3.3.3.2. Convective Brownian diffusion enhancement

The coagulation collision kernel for the Brownian diffusion enhancement $K^{DE}$ is defined by Jacobson et al. (2005) as:

$$K_{i,j}^{DE} = \begin{cases} K_{i,j}^B 0.45 Re_i^{1/3} Sc_{p,j}^{1/3} \\ K_{i,j}^B 0.45 Re_i^{1/2} Sc_{p,j}^{1/3} \end{cases} \quad (5)$$

where $Re_i$ is the Reynolds number of particle i and $Sc_{p,j}$ is the particle Schmidt number of particle j (ratio of viscous to diffusive forces), given in Appendix A. The larger the particle, the higher its fall speed and its Reynolds number, and thus the greater the effect of diffusion enhancement.

### 3.3.3.3. Gravitational collection

The collision kernel for gravitational coagulation is given by:
$$K_{i,j}^G = E_{g\ i,j} \pi (r_i + r_j)^2 |V_{f,i} - V_{f,j}| \quad (6)$$

where $E_{g\ i,j}$ is a collection efficiency term, and $V_{f,i}$ is the terminal fall velocity of particle i (m s$^{-1}$) defined in Appendix A. The collection efficiency $E_{g\ i,j}$ results from the fact that small particles tend to move away from the falling particle, and it is expected to be smaller than unity for most cases. In this paper, we define the collection efficiency (based on the results presented in Jacobson, 2005) as:

$$E_{g\ i,j} = \frac{3y_c^2}{2(r_i + r_j)^2} \quad (7)$$

where $y_c = r_i$ if $r_i \leq r_j$ and $y_c = r_j$ if $r_i > r_j$. Several other equations exist for $E_g$. It was found to be neglected ($E_g$=1, Montmessin, 2002), or lowered by a factor 1/3, 2/3 or 1/6 for water ice (Seinfeld and Pandis, 2006,

Guilbon, 2018, Prappuacher and Klett, 1997), or based on empiric calculations for the terrestrial case (Jacobson, 2005).

### 3.3.3.4. Turbulent inertial motion and turbulent shear

The collision kernel for turbulent inertial motion is given by:

$$K_{i,j}^{TI} = \frac{\pi \varepsilon_d^{3/4}}{g \nu^{1/4}} (r_i + r_j)^2 |V_{f,i} - V_{f,j}| \tag{8}$$

where $\varepsilon_d$ is the rate of dissipation of turbulent kinetic energy per gram of medium (m² s⁻³). A typical value for clear air on Earth is $\varepsilon_d$=0.0005 m² s⁻³ (Pruppacher and Klett 1997), but it can reach up to 0.2 m² s⁻³ during strong convective events. For Mars, we use $\varepsilon_d$=0.15 m² s⁻³ (Martinez et al., 2013). The value is usually larger on Mars than on Earth due to the lower Martian atmospheric density.

The collision kernel for turbulent shear is given by:

$$K_{i,j}^{TS} = \left(\frac{8\pi \varepsilon_d}{15\nu}\right)^{1/2} (r_i + r_j)^3 \tag{9}$$

The characteristic turbulent shear rate at the small length scales is given by the ratio $\varepsilon_d/\nu$. Typical measured values of $(\varepsilon_d/\nu)^{0.5}$ on Earth are on the order of 10 s⁻¹ (Seinfeld and Pandis, 2006), and we expect the same order of magnitude for this ratio on Mars.

### 3.3.3.5. Comparison of the collision kernels

In this section, we compare the collision kernels calculated from the equations of Section 3.3.3 for the different coagulation types of dust particles in terrestrial and Martian conditions, as summarized by Table B-1 (in Appendix B).

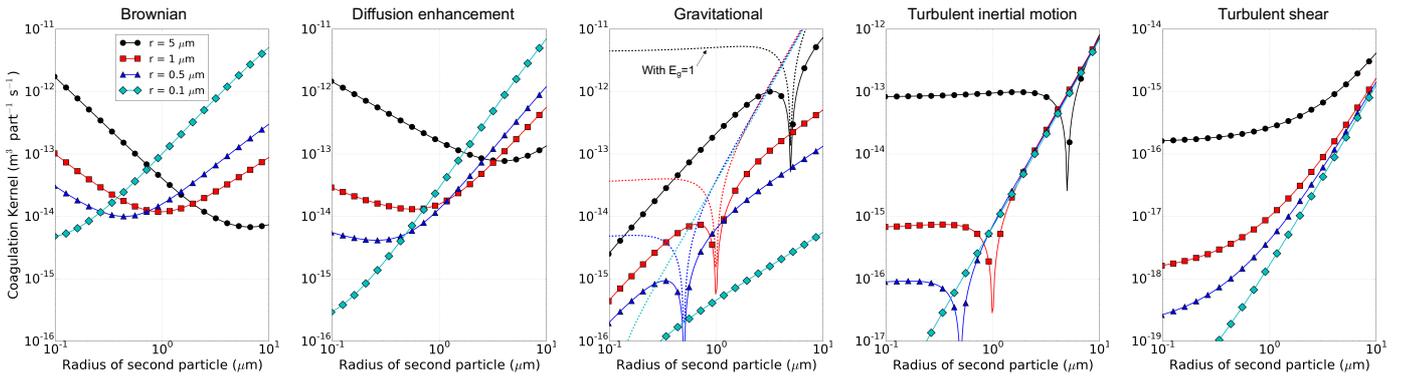

*Figure 1: Collision kernels K (m³ particle⁻¹ s⁻¹) for 4 dust particles of radius 5 µm (black), 1 µm (red), 0.5 µm (blue) and 0.1 µm (cyan) colliding with other dust particles in the range 0.1-10 µm (x-axis) in Mars' atmosphere (typically ~20 km altitude), as calculated from the equations of Section 3.3.3. The main atmospheric parameters are summarized by Table B-1 (in Appendix B). From left to right: Brownian motion, Brownian diffusion enhancement, gravitational, turbulent inertial motion and turbulent shear. The dashed line for gravitational collection corresponds to the case $E_g$=1. Note the change of scale for turbulent inertial motion and turbulent shear. The dips at 0.1, 0.5, 1, and 5 µm for gravitational and turbulent inertial motions result from the fact that the difference in fall velocity (between particles of the same size) is zero at that point.*

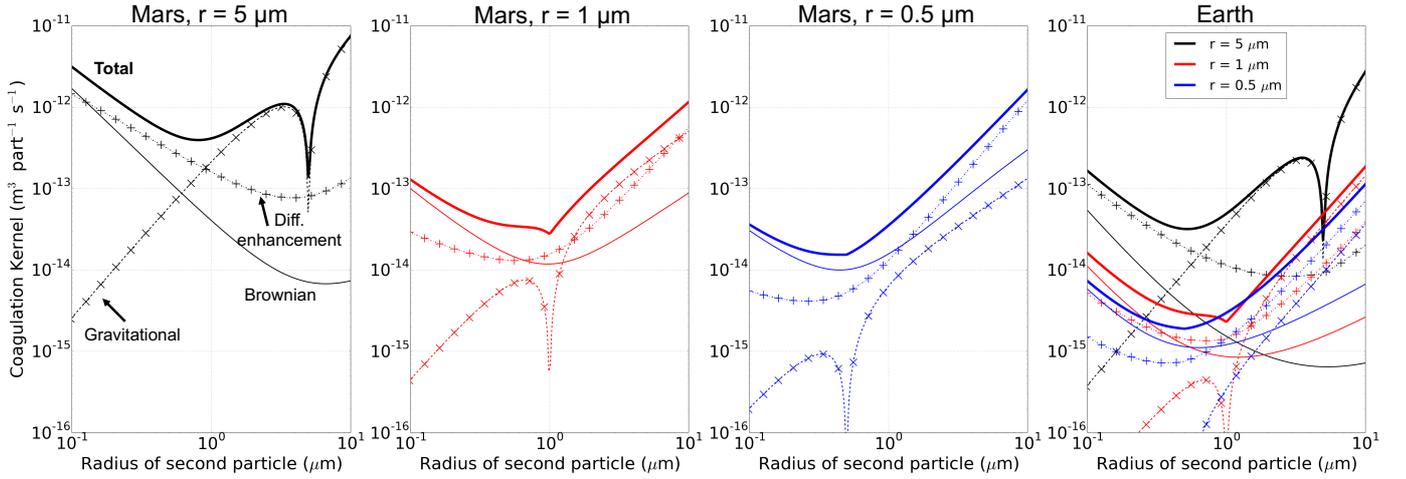

*Figure 2: The 3 main collision kernels ($m^3$ particle$^{-1}$ s$^{-1}$) for 3 dust particles of radius 5 µm (black), 1 µm (red) and 0.5 µm (blue) coagulating with other dust particles in the range 0.1-10 µm (x-axis) in Mars' atmosphere (typically ~20 km altitude, left 3 panels) and Earth's atmosphere (right panel). The thick solid line corresponds to the sum of the three kernels presented in Section 3.3.3 and thus represents the main collision kernel $K=K^B+K^{DE}+K^G$. Brownian collision $K^B$ is shown in a thin solid line, Brownian diffusion-enhanced collision $K^{DE}$ is shown in dashed-dotted line (+), and gravitational collision $K^G$ is shown in dotted line (x). The dips at 0.5, 1 and 5 µm for $K^G$ (dotted lines) result from the fact that the difference in fall velocity (between particles of the same size) is zero at that point.*

Figure 1 compares the collision rates coefficient for dust particles on Mars, and Figure 2 superimposes these coefficients and compares them with those obtained in terrestrial conditions. These figures show that the collision of particles on Mars is dominated by Brownian coagulation for small particles (e.g. 1 µm particles colliding with smaller particles). For larger particles, other types become important, typically Brownian diffusion-enhanced coagulation for collisions of > 5 µm particles with < 1 µm particles and gravitational coagulation, which dominates when only large particles are involved, i.e. for collisions of > 5 µm particles with > 1 µm particles. All three processes play an equal role when only medium-size ~0.5-2 µm particles are involved. In the presence of large particles, the coagulation rate coefficient is large for collisions of large and small particles, which means that small particles would quickly accrete to large particles.

Figure 1 shows that the collision rates from turbulent inertial motion and shear are negligible for dust particles on Mars in most cases: both processes are dominated by Brownian motions so long as at least one particle is small, and by gravitational collection if both particles are large. Consequently, they are neglected on Figure 2 for sake of simplicity, but they are taken into account in the rest of this paper (and in MGCM simulations). We note that very high shear rates would be required on Mars to make the shear collision rate comparable to the Brownian collision rate.

Figure 1 also compares the gravitational coagulation rate coefficient with a case where the collection efficiency $E_g$ between particles i and j (see equation 10) is equal to 1. In this case, the coefficient is higher by at least one order of magnitude and is not sensitive to $r_i$ when $r_j \gg r_i$ (i.e., it does not depend on the radius of the small particle when a collision between a much larger particle is involved). If $E_g$ is not equal to 1, gravitational coagulation can be neglected for submicrometer particles but becomes significant for particles larger than 1 µm.

The Brownian collision rate coefficient is also at a minimum when both particles are of the same size. It rises rapidly when the ratio between the two particle radii increases. This is due to a balance between the particle velocity and surface area: large particles move slowly but have large surface area, which provides a wide target for small, fast particles; small particles move quickly but tend to miss each other because of their smaller surface area.

Figure 2 shows that collision of dust particles is more efficient on Mars than on Earth, assuming the same number density of dust particles. This is due to the higher Knudsen number on Mars, which leads to higher fall velocities and higher diffusion coefficients (see Appendix Figure B-1).

Note that the 150–250 K temperature variation in the Martian lower atmosphere does not have a strong effect on the coagulation rate coefficient and only leads to a maximum change of the collision rate coefficient by a factor of ~1.3 for Brownian motion and Brownian diffusion enhancement, since these coefficients depend on atmospheric temperature as $\sqrt{T}$ (through the thermal velocity equation), which damps the temperature variation. In addition, the Brownian and Brownian diffusion-enhanced coagulation rate coefficients do not depend on the atmospheric density in the free-molecular regime (i.e. for small particles), and therefore these types of coagulation do not depend, to first order, to the altitude (assuming a mixed particle size distribution). The gravitational coagulation rate coefficient depends on the fall velocity $V_f$, which is roughly inversely proportional to the atmospheric density and therefore its effect is increased at higher altitude (assuming the same particle size distribution and number density, i.e. a well-mixed atmosphere).

These results are consistent with those shown in Jacobson et al. (2005) in specific terrestrial conditions and those presented in Montmessin (2002) in Martian conditions.

### 3.3.4. Coalescence efficiency coefficient $E_{coal}$

The equations presented in Section 3.3.3 describe the collision rate of two colliding particles. However, these particles may not always stick together when colliding, as bounce-off could also occur. The probability of sticking is described by the efficiency of coalescence $E_{coal}$. Whereas colliding droplets often merge, the case of solid dust particles on Mars is more complicated. Although it has been argued that the low kinetic energy of two colliding aerosol particles makes bounce-off unlikely (i.e., $E_{coal}$ could be close to unity, Seinfeld and Pandis, 2006), $E_{coal}$ also usually depends on particle shape, composition and other external and more complex mechanisms, such as Van der Waals and viscous forces and electric charging of the particles (see Appendix C). These properties are not well constrained for Martian dust, which makes it difficult to precisely assess the efficiency of coalescence of dust particles on Mars.

For Brownian coagulation, Seinfeld and Pandis (2006) defined a coalescence efficiency $\alpha$ (fraction of collisions that result in coagulation, $\alpha \leq 1$) and adapted their equation 13.56 (equivalent to Equation 4 in this paper) by placing $\alpha$ as a factor of the second denominator term of this equation (i.e., 4 must be replaced by $4\alpha$ in the denominator of Equation 4). However, this definition for $\alpha$ leads to an enhancement of the coagulation rate, so here we argue that $\alpha$ must instead be placed as a divisor of the same denominator term of this equation (i.e., 4 must be replaced by $4/\alpha$ in the denominator of Equation 4). This is consistent with their results: on Earth, in the continuum regime, a coalescence efficiency decreasing from 1 to 0.25 has a small impact of less than 5% on the coagulation rate of a 1 μm particle. In the free molecular regime, the coagulation rate becomes proportional to the coalescence efficiency. This is roughly the case for dust particles on Mars since they are mostly in the free molecular and transition regime, and therefore we assume that $E_{coal}$ is equal to $\alpha$. We also assume that this coefficient is the same for all coagulation types, i.e. $E^B_{coal} = E^{DE}_{coal} = E^G_{coal}$.

The case $E_{coal}=1$ means that particles adhere at every collision. This is the hypothesis made in all previous studies exploring dust particle coagulation on Mars (e.g., Murphy et al., 1990, Montmessin et al., 2002, Fedorova et al., 2014). Although little is known quantitatively about this sticking probability coefficient, some work for terrestrial aerosols seem to indicate that it could be as low as $E_{coal}=0.1$ for submicron and micron-size particles (Phillips et al., 2015, Boulanger, 2017). Consequently, in this study, we will explore both options $E_{coal}=1$ (reference case) and $E_{coal}=0.1$ (low-sticking case, see Section 3.3.5).

### 3.3.5. Summary of general uncertainties

The particle collision rates in the atmosphere, described by the Equations 4-9 in Section 3.3.3, are relatively well validated on Earth and we could expect that they remain valid for the case of dust particles on Mars, although there is a small uncertainty on the collection coefficient term $E_g$ for gravitational collision. On the other hand, many uncertainties remain regarding the coalescence efficiency for dust particles on Mars, which depends on the dust particle properties (material, shape, charges). The combined effect of van der Waals and viscous forces, as well as the fractal shape of particles (see Appendix C and Appendix D), mostly enhance the collision rates, by a factor 1.2-10. The net effect of electrical charging of the particles (Coulomb forces) remains unclear.

In addition, when several coagulation types are significant (e.g., both the Brownian and gravitational coagulations), the coagulation kernels (e.g. $\beta^B$ and $\beta^G$) are commonly added to predict the behavior of the particles. However, this simple addition, suggesting that all coagulation types act independently, does not apply in practice and the combined kernel was observed to be less by a correction factor $\omega_s$ up to 1.5 (Simons et al., 1986) in specific terrestrial conditions.

In this context, we will explore the coagulation of dust particles on Mars following different scenarios. We will assume:
- A reference case, with $E_{coal} = 1$ (i.e. $\beta=\Sigma K$), which means that the coalescence efficiency, the external effects (van der Waals, viscous, Coulomb forces), the fractal case, and the correction factor $\omega_s$ balance each other for a null net impact on dust coagulation, independently of the particle size. This case will be explored with the reference equation for $E_g$ (Equation 7) and with $E_g=1$ and $E_g=0$ (no gravitational coagulation).
- A low-sticking scenario, with $E_{coal} = 0.1$ (i.e. $\beta=0.1\Sigma K$), which corresponds to a case in which the mechanisms leading to a diminution of the coagulation rate (the coalescence efficiency, the Coulomb and viscous forces, the correction factor $\omega_s$) overcome those leading to an enhancement (van der Waals forces, fractal case).

## 3.4. Timescales for coagulation processes

Since the coagulation of particles depends on the particle number concentration n(r) and the size distribution r, one must know the initial particle population to assess the impact of coagulation and estimate the related timescale. Montmessin (2002) defined the characteristic time for coagulation $\tau_{coag}$ as the time needed for a particle i to double its volume by coagulation with particles j, which is given by an integration of the coagulation kernels over the particle size distribution:

$$\tau_{coag}^{-1} = \int_0^\infty \delta_{ij} \frac{V_j}{V_i} \beta_{ij}^{B+G+DE} n_j dr_j + \int_0^\infty (1-\delta_{ij}) \beta_{ij}^{B+G+DE} n_j dr_j$$

$$\delta_{ij} = \begin{cases} 1 \text{ if } r_j < r_i \\ 0 \text{ if } r_j > r_i \end{cases} \quad (10)$$

Where $V_j/V_i$ is the volume ratio of particle j by particle i. As shown by this equation, the timescale for coagulation is shorter (i.e., coagulation is more efficient) for higher particle number concentrations and larger particle size spectra (higher variance).

Figure 3 shows the timescale for coagulation for the different particle sizes of a lognormal distribution (represented by the geometric cross-section weighted mean particle radius, or effective radius $r_{eff}$ and the effective width of the distribution, or variance, $\sigma_{eff}$), corresponding to Martian non-storm conditions (particle density N=6 cm$^{-3}$, $r_{eff}$=1.5 µm, $\sigma_{eff}$=0.5). Brownian coagulation dominates for small particles, while gravitational coagulation dominates for large particles (and turbulent inertial motion for very large particles > 30 µm), as discussed in Section 3.4 and as also shown by Figure 2. Coagulation through turbulent shear is negligible, as its timescale is greater than $10^{10}$ s for all particle sizes.

Small particles quickly accrete to large ones: the lifetime of a 0.01 µm particle is about one day, and the lifetime of particles smaller than 0.1 µm is less than 3 weeks. For most of the dust particles in these distributions, whose radius is around 0.5-2 µm, the timescale for dust coagulation is about $10^7$ s, i.e., ~4 months, which is longer than the typical duration of regional dust storms but is about the duration of global dust storms. For the sake of comparison, the timescale for gravitational coagulation assuming a collision efficiency $E_g$=1 (maximum efficiency, which is likely overestimated) is also shown on Figure 3. In this case, gravitational coagulation would dominate for particle radii greater than 0.2 µm and would keep the coagulation timescale below ~60 days.

Figure 4 shows how the timescale for coagulation is impacted for different particle size distributions. As expected, the coagulation timescale is longer for the distribution with the lower variance (red line) and shorter for the distribution with the larger effective radius (orange line). The timescale is inversely proportional to the particle number density and is therefore 10x shorter if the particle number density is multiplied by 10. Figure 4 shows a case with N=60 cm$^{-3}$, which would correspond to dust storm conditions. In this case, the lifetime of a 1 µm particle is about 12 days.

However, as mentioned in Section 1, the efficiency of coagulation can also be enhanced by convective movements (e.g., the ascending cell of the Hadley circulation) and vertical mixing that allow for the particles to be suspended in the atmosphere for longer periods.

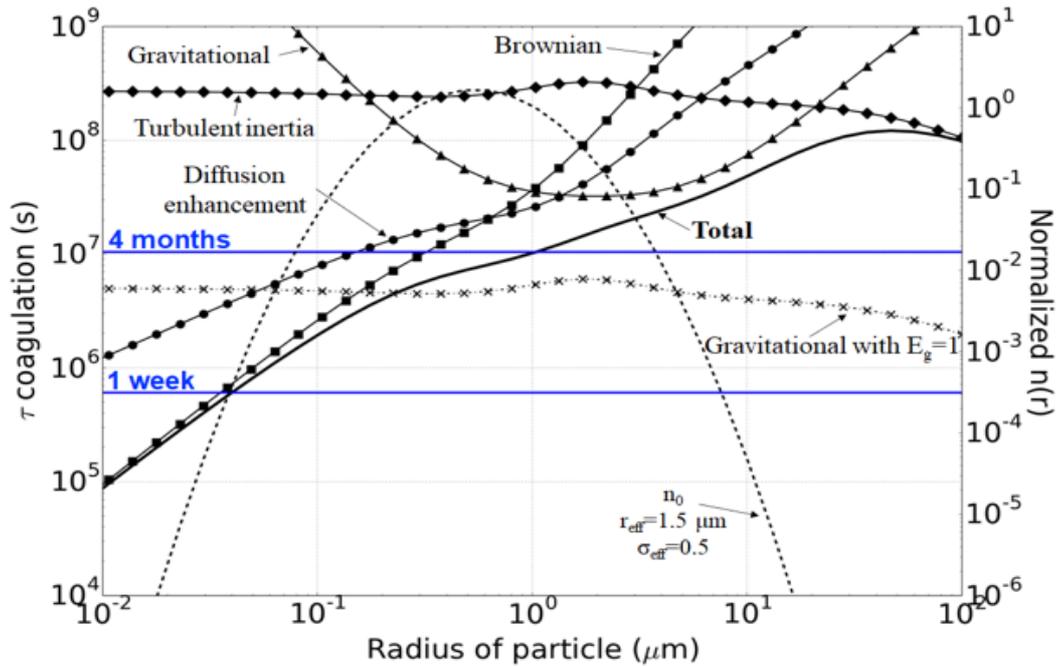

*Figure 3: Coagulation timescale $\tau_{coag}$ (thick black line) for dust particles on Mars versus their particle size for one particle size distribution (dashed line). The contribution of each coagulation type is shown (Brownian motion: squares; Brownian diffusion enhancement: circles; Gravitational: triangles; Turbulent inertial*

motion: diamonds). The timescale for gravitational coagulation assuming a sticking efficiency $E_g=1$ is also shown (dash-dotted line) but does not account for the total coagulation timescale. The corresponding atmospheric lognormal particle size distribution $n_0$ is superimposed (dashed lines, right y-axis) and correspond to typical Martian conditions, with a total particle number density of $N=6$ cm$^{-3}$ (air density $\rho_a=0.002$ kg m$^{-3}$, number mixing ratio $3\times10^9$ particles kg$^{-1}$), an effective radius $r_{eff}=1.5$ µm and an effective radiance $\sigma_{eff}=0.5$. The distribution is normalized to their particle number density at 1 µm.

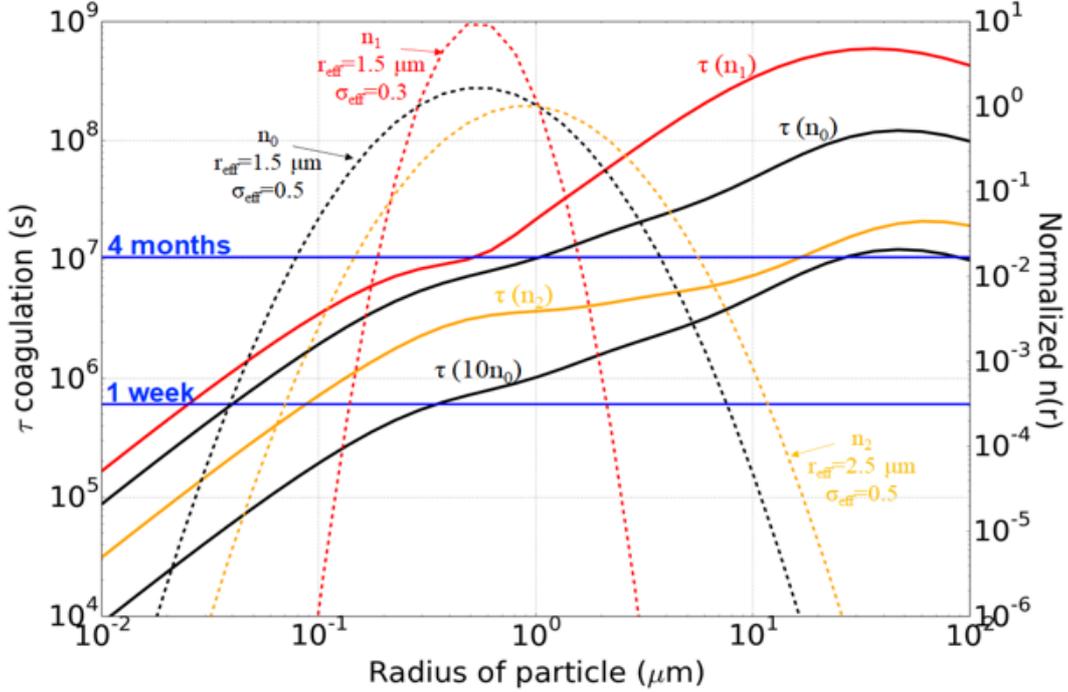

*Figure 4: Coagulation timescale $\tau_{coag}$ (thick lines, left y-axis) for dust particles on Mars versus their particle size for different particle size distribution (dashed lines, superimposed, right y-axis). The distributions are normalized to their particle number density at 1 µm. Black: reference case with $N=6$ cm$^{-3}$, $r_{eff}=1.5$ µm, $\sigma_{eff}=0.5$ as in Figure 3. A case with 10x more particles ($N=60$ cm$^{-3}$) is shown. In this case, the shape of the distribution remains the same due to the normalization but the timescale for coagulation is increased by a factor of 10. Red: Same as reference case but with $\sigma_{eff}=0.3$. Orange: Same as reference case but with $r_{eff}=2.5$ µm.*

# 4. Simulations results of dust coagulation in non-storm and storm conditions

In this Section, we first present the 3-D NASA Ames Mars Global Climate Model (MGCM) used in this study to parameterize dust coagulation processes (Section 4.1). An important point is the fact that the atmospheric dust in this model is always represented by a lognormal particle size distribution, with two moments, namely the effective radius $r_{eff}$ and the effective variance of the distribution $\sigma_{eff}$, the latter remaining fixed over time and space in the model. Several physical processes are parameterized in the MGCM with this assumption that the particle size distribution remains lognormal (e.g., dust lifting, dust sedimentation, radiative transfer). Regarding coagulation, Friedlander and Wang (1966), Rossow (1978) and Montmessin (2002) stated that the normalized size distribution evolves towards the same invariant shape with constant variance regardless of the initial conditions, while the total number density and mean particle mass continue to change with time. However, this has not been well quantified. In Appendix E, we verify with 0-D numerical tests of the coagulation equations that an initially lognormal particle size distribution can still be approximated as a lognormal distribution after it evolves through coagulation processes, and we quantify the error made with this assumption. Finally, we present the simulations of the 2018 global dust storm with the MGCM and active coagulation in Section 4.2.

# 4.1. The NASA Ames Mars GCM

## 4.1.1. General settings

We use the NASA Ames Mars GCM (MGCM), which employs the NOAA/GFDL cubed-sphere finite-volume (FV3) dynamical core and physics packages from the Ames Legacy MGCM as described in Kahre et al. (2018) and Haberle et al. (2019). In particular, the MGCM includes topography from the Mars Orbiter Laser Altimeter (MOLA), albedo and thermal inertia maps derived from Viking and Mars Global Surveyor (MGS) Thermal Emission Spectrometer (TES) observations, coupled dust and water cycles (as described in Haberle et al., 2019 and Bertrand et al., 2020), water ice cloud microphysics (Montmessin et al., 2002, 2004, Nelli et al., 2009, Navarro et al., 2014), a planetary boundary layer scheme (Haberle et al., 2019 and references therein), and a 2-stream radiative transfer scheme that accounts for gaseous absorption of $CO_2$, $H_2O$ and scattering aerosols, including dust and water ice particles (Toon et al., 1989, Haberle et al., 2019), with optical properties of both dust and water ice cloud particles that depend on particle size and thus evolving in time and space. More details about the model are also given in Bertrand et al. (2020).

The lognormal particle size distributions of dust and clouds are represented by a two-moments scheme, with a spatially and temporally varying particle mass and number mixing ratios (thus a varying effective particle radius), and a constant effective variance. Dust is lifted following an "assimilated dust lifting" scheme (e.g. Kahre et al., 2009; Greybush et al., 2012) based on the observed column opacity fields (from MCS for Mars Year 34), as in Bertrand et al. (2020; see their section 3.3.2; see also Kahre et al., 2009; Greybush et al., 2012). The lifted dust particle size distribution is assumed to be lognormal. Bertrand et al., 2020 used an effective lifted particle radius $r_{eff}=3$ μm and an effective variance $\sigma_{eff}=0.5$. Here we will use this effective lifted particle radius for our baseline MGCM simulation but also explore smaller radii with $r_{eff}=2$ μm and $r_{eff}=1$, which is a more extreme case (the variance remains unchanged in all simulations).

As in Bertrand et al. (2020), the MGCM simulations described in this paper are carried out with a horizontal resolution of 2°x2° and 46 vertical levels, with a vertical resolution decreasing from 20 m near the surface to 10 km at the model top (~80 km). The simulations are carried out with radiatively active dust over the global dust storm of Martian Year 34 using the same initial state and settings as in Bertrand et al. (2020), and one spin-up year. One notable exception is that in this study, we do not include the radiative and sedimentation effects of water ice clouds (deposition of ice on dust is turned off). The impact of water ice clouds appears to be minimal during the global storm, and this allows us to focus on the coagulation effect in a dust-only simulation.

## 4.1.2. Implementation of coagulation equations in the MGCM

Coagulation of dust particles is parameterized in the MGCM by implementing the equations presented in Section 3. The coagulation kernels $\beta_{i,j}$ are interpolated from values stored in a lookup table covering a wide range of particle radii ($N_B=40$). This allows for a reduction in computing time by a factor of 10.

Because the MGCM assumes lognormal aerosol distributions in many physical parameterizations (e.g., in the radiative transfer and sedimentation routines), the model is forced to keep a lognormal atmospheric particle size distribution and a constant effective variance during the coagulation process. We estimate the error made when approximating the final distribution after coagulation as lognormal in Appendix E.

In each atmospheric layer of each model grid, and at each physical timestep h, the coagulation routine of the MGCM reads the atmospheric state (temperature, density) and the local dust particle mass and number mixing ratio, computes the equations of coagulation and returns a loss rate of the number mixing ratio (the mass is conserved). The steps of the algorithm are the following:

1. **Initial distribution:** the model assumes that the dust population follows a lognormal distribution with a fixed variance $\sigma_{eff}$=0.5. From the input dust mass $M_0$ (kg$_{dust}$ kg$^{-1}$) and number $N_0$ (particle kg$^{-1}$) mixing ratio, it computes the mean radius of the lognormal distribution $r_0$:

$$r_0 = \left[\frac{3}{4}\left(\frac{M_0}{\pi N_0 \rho_d}\right)\right]^{1/3} exp(-1.5\sigma_0^2) \tag{11}$$

Where $\rho_d$ is the particle density (set to 2500 kg m$^{-3}$). In the 0-D simulations of Section Appendix E, the initial distribution is described by the effective radius $r_{eff}$ and variance $\sigma_{eff}$ and:

$$\begin{aligned} r_0 &= r_{eff}/(1+\sigma_{eff})^{5/2} \\ \sigma_0^2 &= \ln(1+\sigma_{eff}) \end{aligned} \tag{12}$$

2. **Discretization:** the initial particle density n, at timestep t, is discretized over $N_B$=40 particle radii $r_i$ (particles bins) and follows a lognormal expression:

$$n(r_i) = \frac{N_0 \rho_a \Delta d_i}{2r_i\sqrt{2\pi}\sigma_0} exp\left[-\frac{\ln^2\left(\frac{r_i}{r_0}\right)}{2\sigma_0^2}\right] \tag{13}$$

Where $\rho_a$ is the air density (kg m$^{-3}$) and $\Delta d_i$ is the diameter width of the lognormal distribution, given by:

$$\Delta d_i = 2r_i 2^{1/3} \frac{V_{rat}^{1/3}-1}{(1+V_{rat})^{1/3}} \tag{14}$$

Where $V_{rat}$ is the constant volume ratio given by equation 1.

3. **Coagulation kernels:** the coagulation kernels are computed from the equations of Section 3.3.3.

4. **Coagulation and change in particle density n:** The new particle density n(r) at timestep t+h is calculated from n(r) at timestep t using equation 3. The timestep h corresponds to a physical timestep of the MGCM. This step could be a loop in which n evolves through coagulation until the final timestep. However, in the MGCM, we force the particle size distribution to be lognormal, with a constant variance (imposed by the model design), hence the following steps.

5. **Forcing to a lognormal distribution with constant variance:** the number mixing ratio of the new distribution at timestep t+h is first calculated from the particle density n(r) at t+h:

$$N_{t+h} = \sum_{i=0}^{N_B} \frac{n(r_i)}{\rho_a} \tag{15}$$

Then the model assumes that the new particle size distribution is lognormal and computes a new mean radius $r_0$ from the mean variance and mass mixing ratio (unchanged) and the new number mixing ratio $N_{t+h}$, using equation 11. Finally, the particle density n(r) at t+h is rewritten as following a lognormal distribution, using equation 13 with the new mean radius $r_0$ and number mixing ratio $N_{t+h}$.

6. **Conservation of volume:** conservation of volume is ensured with the following:

$$n_{t+h}(r_i) = n_t(r_i) \frac{\sum_{i=0}^{N_B} n_t(r_i) V_i}{\sum_{i=0}^{N_B} n_{t+h}(r_i) V_i} \qquad (16)$$

The algorithm then loops over these steps 1-6. The loss rate in particle mixing ratio $N_{t-h} - N_t$ is output by the model at each physical timestep h.

## 4.2. Coagulation during the 2018 global dust storm simulated with the 3-D NASA Ames Mars global climate model

In Section 4.2.1, we present our reference simulation results obtained with the MGCM during the full storm, as in Bertrand et al. (2020), but with active coagulation. In Section 4.2.2, we present simulation results of the decay phase of the storm only, in which dust lifting is shut off. This allows us to assess the net effect of the different types of coagulation during the storm's decay without any surface source of dust involved (other active processes only include dust transport and sedimentation).

### 4.2.1. Simulation results of the global storm with active coagulation

Figure 5 shows the tropical mean IR column dust opacity as observed by MCS and as modeled with the MGCM, with and without coagulation. For all simulations, the MGCM opacities remain similar and in good agreement with observations before the storm and during its onset. Then, discrepancies are obtained during the storm's decay. During this phase, the column dust opacities decline quasi-exponentially with time returning to their background levels ~90 sols after the storm's peak. This phase is strongly controlled by dust sedimentation, and therefore is very sensitive to the dust particle size (smaller particles lead to smaller sedimentation rates and a subsequent slower storm decay).

The reference simulations with a lifted effective radius $r_{eff}$=3 μm and $r_{eff}$=2 μm produce a decay in opacity slightly slower than that observed, with coagulation improving the match. On the other hand, the reference simulation with a lifted effective radius $r_{eff}$=1 μm but without coagulation produces a very slow decay in opacity, far from what is observed. When coagulation is active, small particles are accreted to large particles, thus the mean particle radius increases and the decay rate increases. As a result, despite their differing lifted effective radii, all three simulations with active coagulation produce a similar decay phase in opacity, close to that observed.

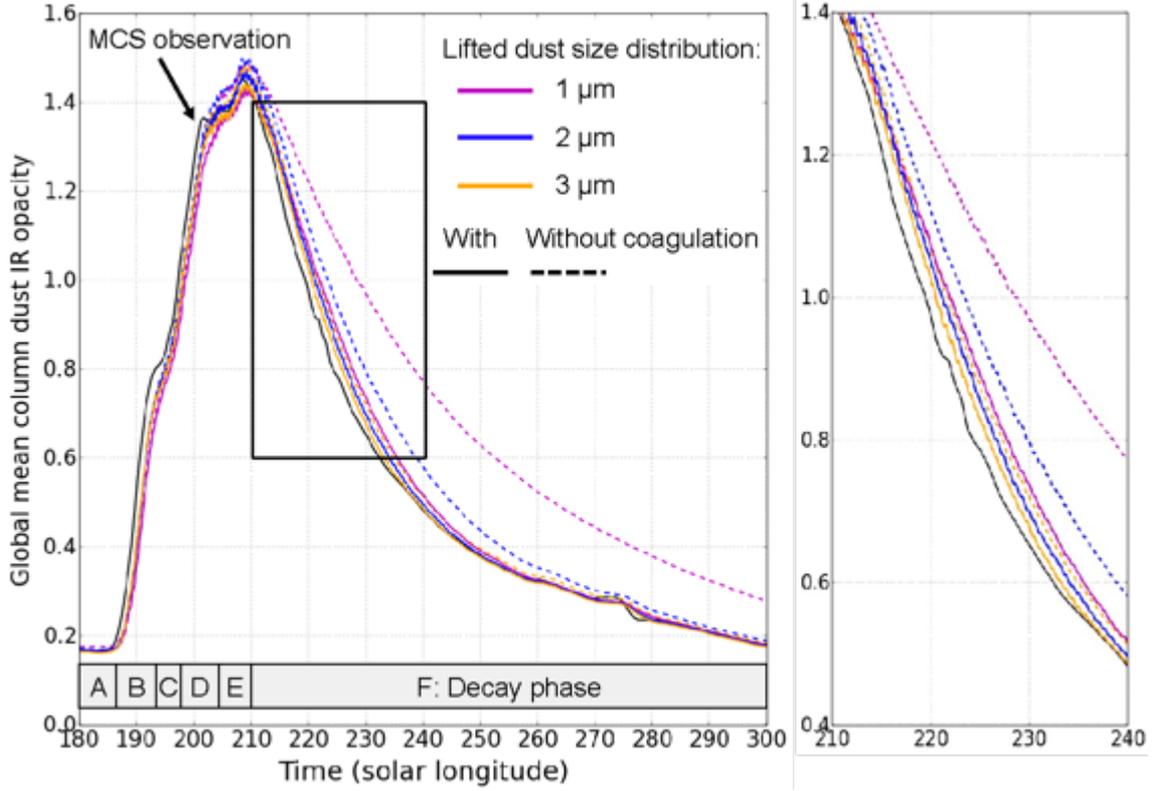

*Figure 5: Column dust IR (9.3 μm) opacity (zonal mean, 50°S-50°N) during the global storm as observed by MCS (black, Montabone et al., 2020) and as modeled with the MGCM assuming an effective lifted dust particle size of 1 μm, 2 μm and 3 μm, with (solid) and without (dashed) active coagulation. The different phases of the storm are indicated as in Bertrand et al., 2020: pre-storm (A), onset (B; near Acidalia/Chryse Planitia, Ls=187°-193°), expansion (C; eastward and southward, Ls=193°-196°), Tharsis extreme events (D; intense lifting, high column dust opacity, dust plumes or "tower" near Tharsis region, with strong ramping up of the dust heating, Ls=196°-204°, Heavens et al., 2019), mature phase (E; maximum atmospheric dust loading and temperatures) and decay phase (F).*

During the peak and decay phase, coagulation increases the mean effective radius of dust in the atmosphere by almost a factor of 2 in the case $r_{eff}$=1 μm (up to +1 μm), and by a factor of ~1.3 in the case $r_{eff}$=3 μm (up to +0.5 μm), as shown by Figure 6. At the end of the decay (Ls=300°), dust particles are significantly larger than the pre-storm period in the case $r_{eff}$=1 μm.

Figure 7 shows the dust number density obtained in our simulations with and without coagulation and the column mean coagulation rates in particle cm$^{-2}$ s$^{-1}$, while Figure 8 shows the coagulation rates in the atmosphere at Ls=180°, 210° and 300°. In the case $r_{eff}$=1 μm, the number density is strongly impacted by coagulation as it removes significant amounts of small particles. During the pre-storm period, column mean coagulation rates are about $10^9$ particles cm$^{-2}$ s$^{-1}$, and coagulation reduces the number density by a factor of 2 at all altitudes, compared to the simulation without active coagulation. During the storm, the coagulation rates are up to 10 times larger (maximum at storm's peak, Ls=210°), and the number density is reduced by a factor of up to 10, compared to the simulation without active coagulation. In the case $r_{eff}$=3 μm, coagulation has little effect during the pre-storm period (coagulation rates ~$2×10^7$ particles cm$^{-2}$ s$^{-1}$), but reduces the number density by a factor up to 2 during the storm (coagulation rates ~$4×10^8$ particles cm$^{-2}$ s$^{-1}$).

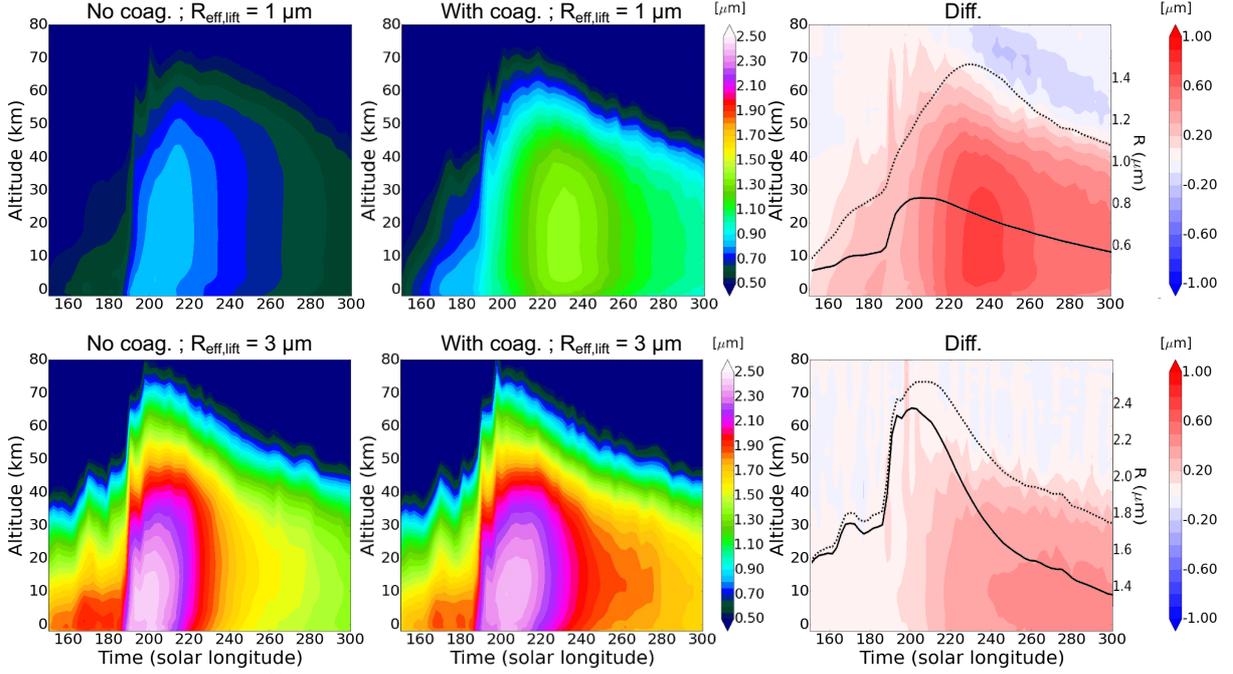

*Figure 6: Dust particle effective radius (zonal mean, 30°S-30°N) during the storm as obtained with the MGCM for simulations without (left) and with (middle) active coagulation, and the differences (right), assuming an effective lifted dust particle size of 1 μm (top) and 3 μm (bottom). The superimposed black line on the right panel indicates the column-averaged effective radius (calculated from the total column dust mass and number density) obtained with (dashed line) and without (solid line) coagulation.*

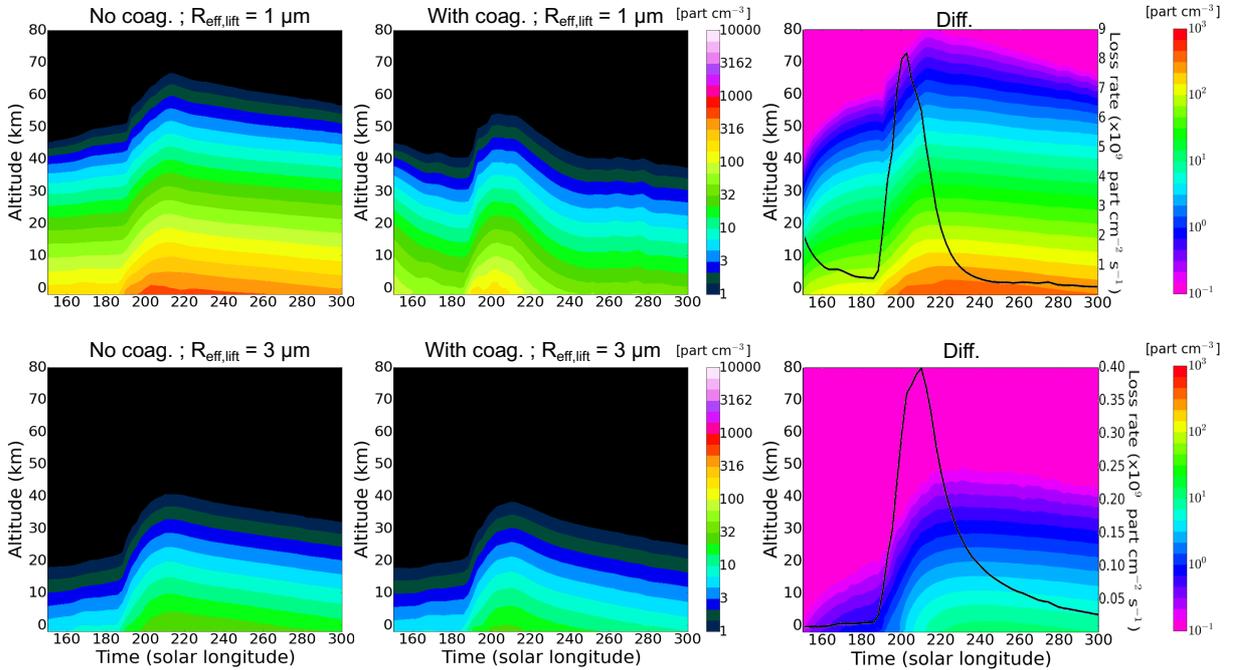

*Figure 7: Dust number density (particle cm$^{-3}$, zonal mean, 30°S-30°N) during the storm as obtained with the MGCM for simulations without (left) and with (middle) active coagulation, and the difference between both (right), assuming an effective lifted dust particle size of 1 μm (top) and 3 μm (bottom). The superimposed black line indicates the loss rate of particles in the column due to coagulation (particle cm$^{-2}$ s$^{-1}$).*

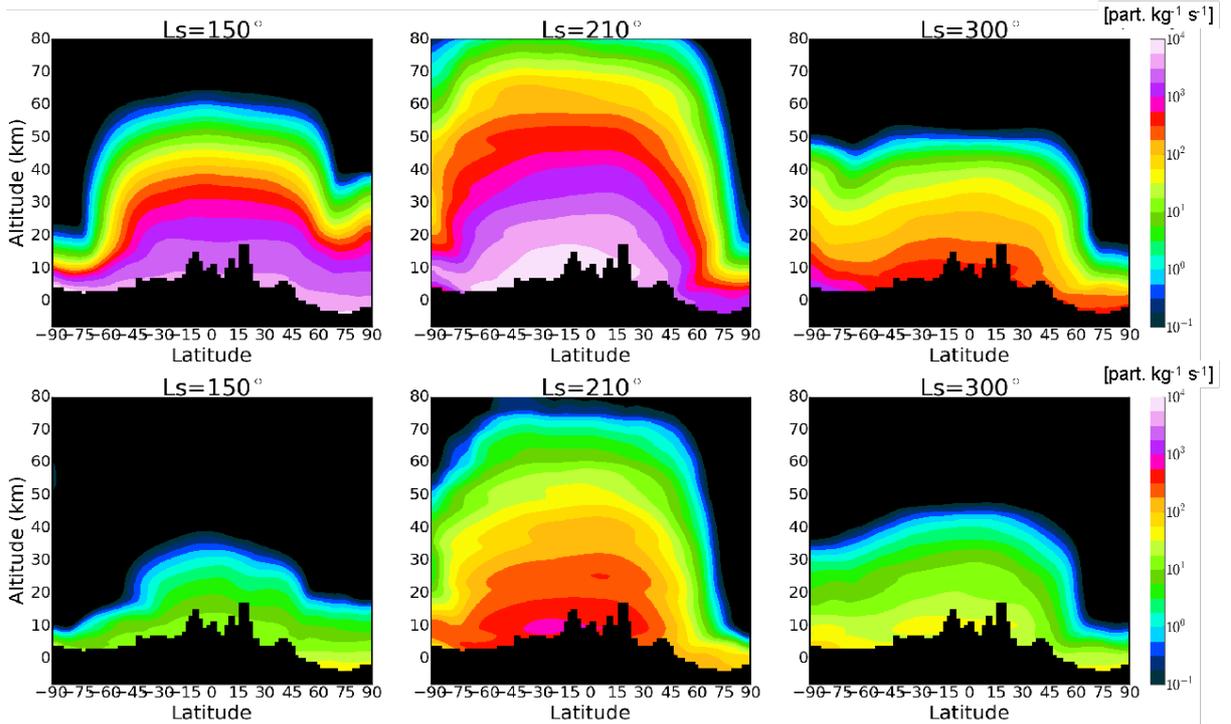

*Figure 8: Coagulation rate (particle kg$^{-1}$ s$^{-1}$) averaged over 4 sols around Ls=150° (pre-storm, left), 210° (storm's peak, middle), and 300° (end of decay), as obtained with the MGCM with coagulation, assuming an effective lifted dust particle size of 1 μm (top) and 3 μm (bottom).*

Figure 9 and Figure 10 highlight the impact of coagulation on the climate (vertical distribution of dust, general circulation, temperature and zonal winds) at Ls=210° (storm's peak) while Figure 11 and Figure 12 show similar results at Ls=300° (end of decay phase), as simulated by our GCM. These figures show that coagulation has a significant impact on the climate in the $r_{eff,lift}$=1 μm case, and a limited impact in the $r_{eff,lift}$=3 μm case. In the $r_{eff,lift}$=1 μm case, large numbers of small particles are removed in the upper atmosphere (by enhanced sedimentation, following coagulation, Figure 9, top), leading to an extent of the dust 10 km lower (Figure 9, Figure 10, top), a 10-20 K colder atmosphere above 40 km altitude at Ls=210° (Figure 11, top), and above 10 km altitude at Ls=300° (Figure 12, top). At Ls=210° during storm's peak, in the first 20 km kilometers above the surface, warmer atmospheric temperatures are obtained with coagulation due to the presence of larger "coagulated" dust particles and a larger mass of dust (Figure 11, top), which increase the radiative daytime warming and nighttime cooling of the atmosphere, and hence increase the strength of the general circulation (Figure 9, top). At Ls=300°, i.e. at the end of the decay phase, coagulation and subsequent sedimentation strongly reduced the amount of dust in the atmosphere. The dust-top altitude as well as the branches of the Hadley cell circulation are less deep in the upper atmosphere compared to the case without coagulation. The zonal flow reflects the behavior of the meridional circulation, with equatorial eastward and polar westward jets located at lower altitudes compared to the case without coagulation. Finally, these figures also highlight the fact that the atmospheric state simulated with $r_{eff,lift}$=1 μm case and with coagulation converges toward that obtained with $r_{eff,lift}$=3 μm, in particular at the end of the decay phase due to coagulation constantly increasing the effective particle size during the storm.

In the case $r_{eff}$=3 μm, the impact of coagulation is less as there are fewer small particles to be accreted and the coagulation rates are slower. The vertical dust distribution and general circulation remain relatively unchanged after the storm decay (Figure 9, Figure 10, bottom). The larger differences are obtained at Ls=300°, with temperatures up to 5 K colder in the atmosphere and equatorial zonal winds 5 m s$^{-1}$ slower at 40 km altitude with coagulation compared to the case without coagulation (Figure 11, Figure 12, bottom).

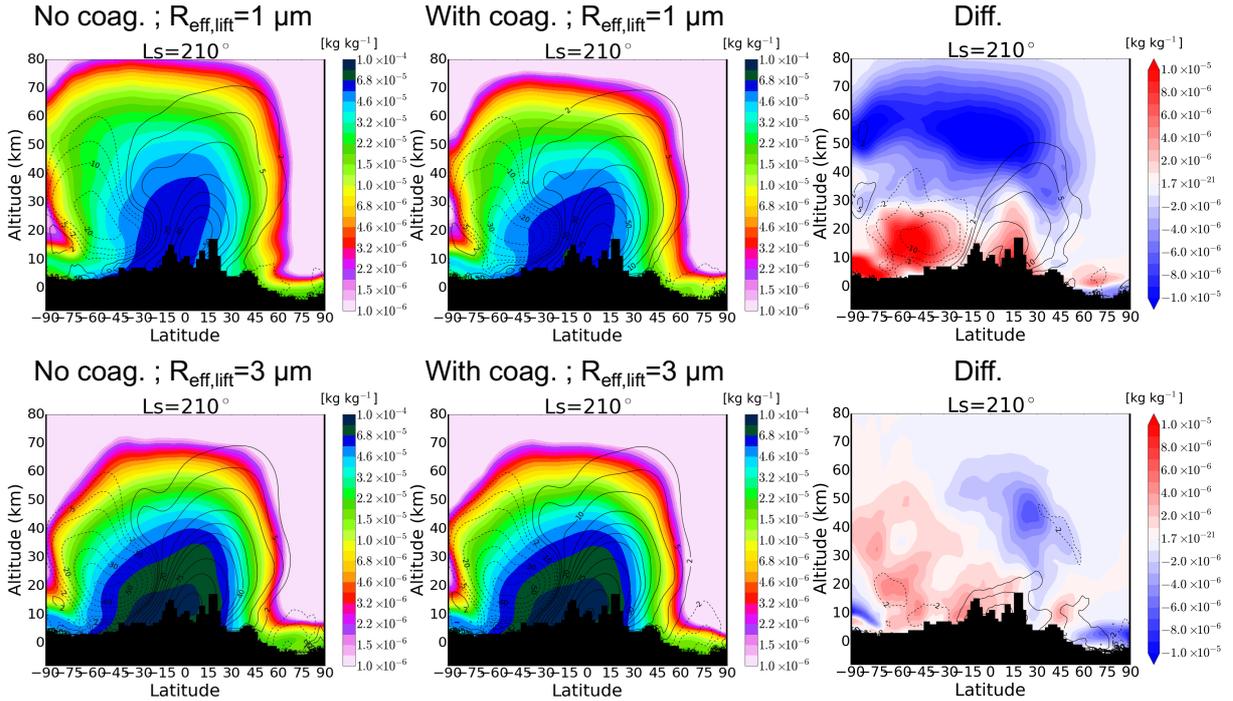

*Figure 9: Difference in zonal mean atmospheric dust mass mixing ratio (kg kg$^{-1}$) averaged over 4 sols (filled contours) around 210° (storm's peak), as obtained with the MGCM with coagulation (left) and without (middle) assuming an effective lifted dust particle size of 1 µm (top) and 3 µm (bottom). The zonal mean streamfunction ($10^8$ kg s$^{-1}$) is shown as black contour lines, with positive values (solid black lines) indicating clockwise circulation. Differences (with coagulation minus without) are shown on the right panels.*

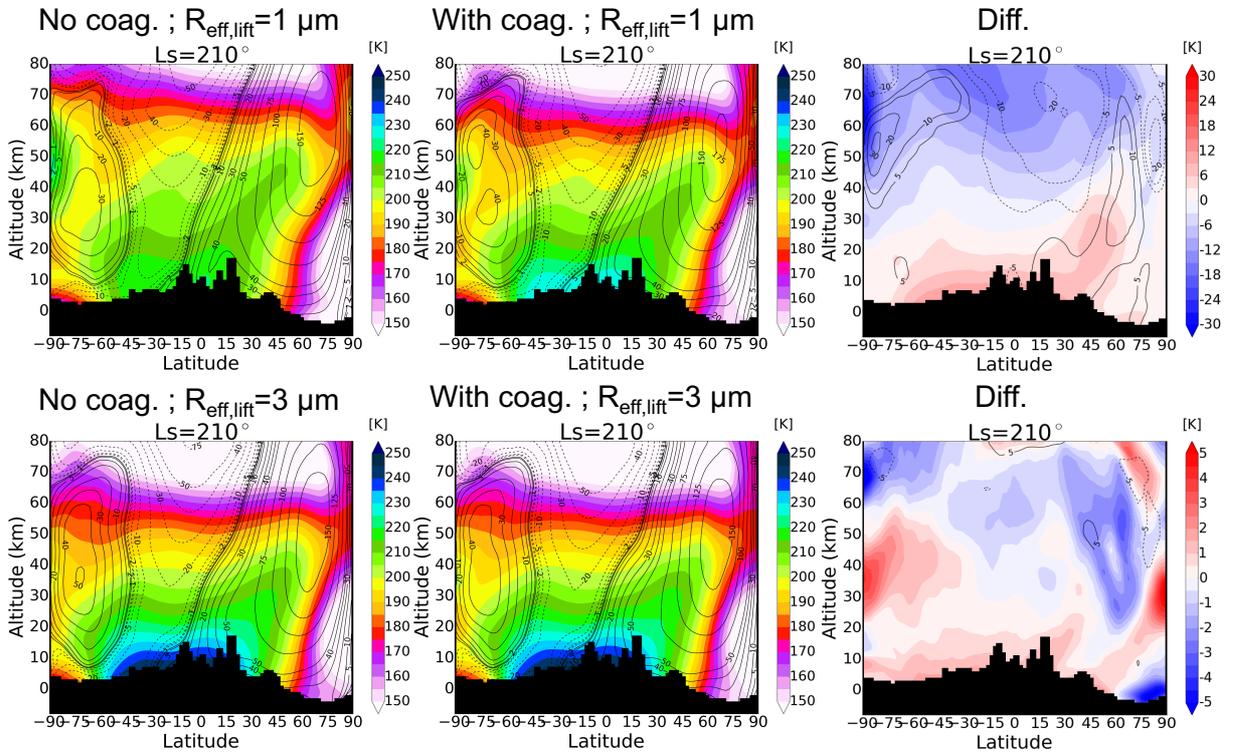

*Figure 10: As Figure 9 but showing the zonal mean atmospheric temperatures, with zonal mean zonal winds as black contour lines. Note the changed color bar for the 3 µm simulation results in the bottom right panel.*

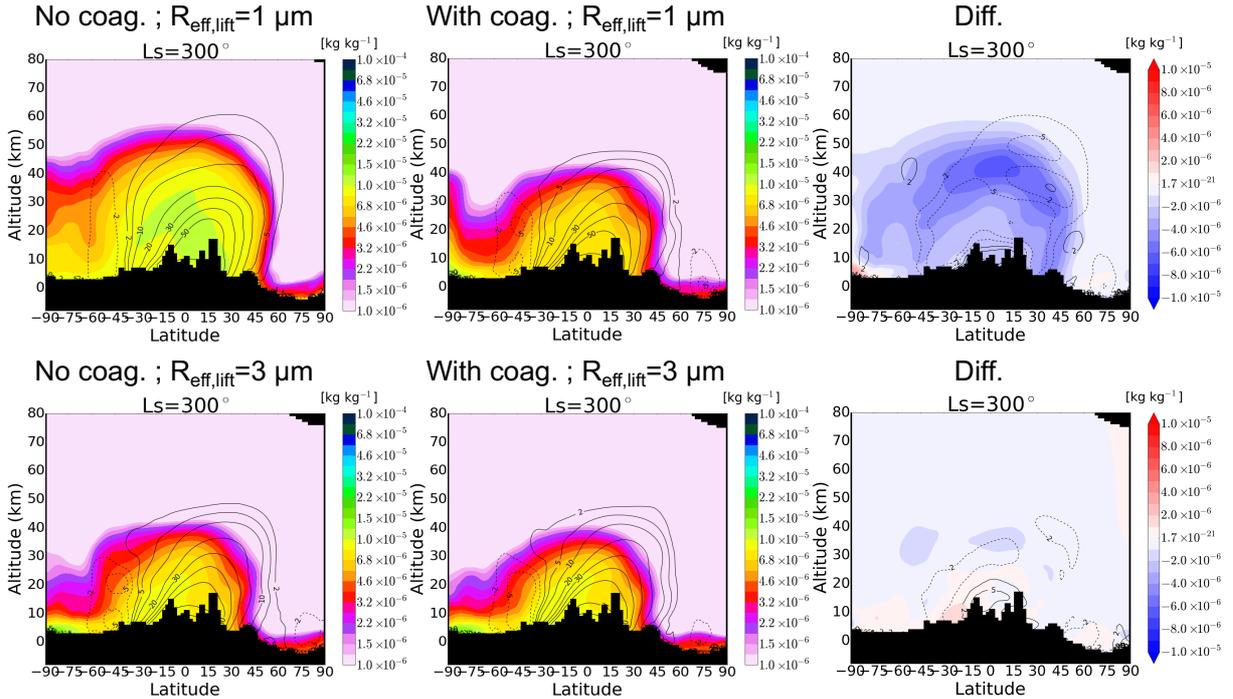

*Figure 11: As Figure 9, but at Ls=300° (end of decay).*

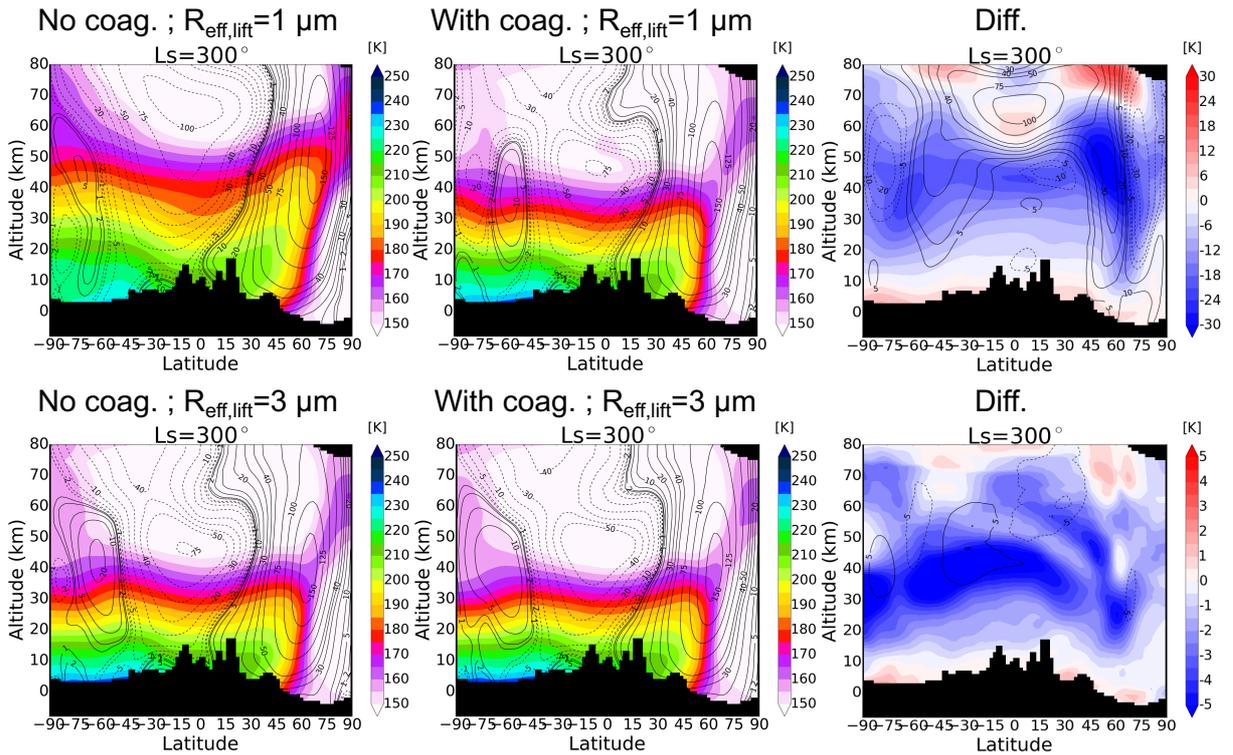

*Figure 12: As Figure 10, but at Ls=300° (end of decay).*

Outside of the storm (i.e., before Ls=188° in MY34) changes in dust distribution, general circulation, temperature and winds are negligible in case $r_{eff,lift}$=3 μm (not shown, changes are less than 1%), but remain significant in the case $r_{eff,lift}$=1 μm. This is highlighted by Figure 13, showing the zonal mean atmospheric temperatures and zonal mean zonal winds at Ls=150°, and Figure 14, showing the atmospheric mean effective radius of dust during northern spring and summer as obtained in the GCM with and without coagulation, in the case $r_{eff,lift}$=1 μm. With coagulation, dust particle size stabilizes around 0.8-0.9 μm, versus 0.4-0.5 µm without coagulation. A ~10 km lower dust-top altitude is obtained, leading to a ~15 K colder atmosphere and changes in zonal winds of up to 20 m s$^{-1}$ at 40 km, compared to the case without coagulation. Note that these simulations are performed without water ice clouds, which strongly impact the atmospheric state at this season.

Finally, our results show that coagulation of dust, if it actually occurs in the Martian atmosphere, would be a significant process affecting the dust cycle and the climate, providing that large amounts of submicron particles are lifted from the surface. The implications of these results are further discussed in Section 5.

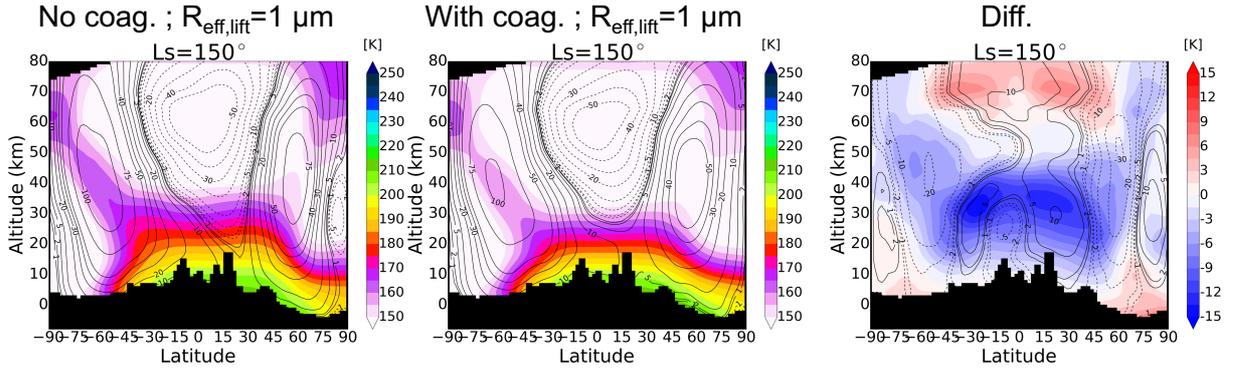

*Figure 13: As Figure 10, but at Ls=150° (pre-storm period) for $r_{eff,lift}$ = 1 µm.*

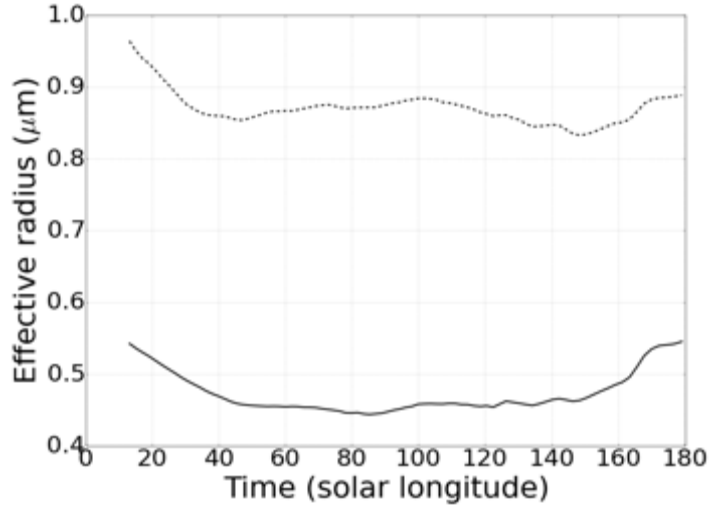

*Figure 14: Mean effective radius of dust particle in the atmosphere during northern spring and summer (no dust storm) averaged between 50°S-50°N in our MGCM simulation performed with a lifted radius $r_{eff,lift}$=1 µm with (dashed line) and without (solid line) active coagulation.*

### 4.2.2. Simulation of the decay phase of the storm with active coagulation (no dust lifting)

In an alternative MGCM simulation, dust lifting is shut off at Ls=210° (storm's peak), and we explore the effects of the different types of coagulation during the storm's decay. Figure 15 shows the decay in IR opacity obtained when only Brownian coagulation is active, and when Brownian diffusion-enhanced and gravitational coagulation are also active. We also show the evolution of opacity when the coalescence factor is 0.1 or 0.5 (ie. when the coagulation is 10x and 2x less efficient, respectively). The coagulation rates at 20 km altitude are shown for these different scenarios on Figure 16.

In the simulations performed with a lifted dust particle size of $r_{eff,lift}$=1 µm, coagulation is dominated by Brownian motion, with Brownian diffusion enhancement having a significant impact as well by increasing the Brownian coagulation rates by 30%, for a total of 24,000 coagulated particles per kg of air per second at 20 km altitude at storm's peak. Gravitational coagulation is negligible in this case. The simulations performed with a coalescence factor of 0.1 still produces a slight but non-negligible decrease in opacity. In the simulations performed with a lifted dust particle size of $r_{eff,lift}$=3 μm (ie. larger particles are injected), the coagulation rates are much less, decreased by a factor 100. Brownian motion still dominates, but Brownian diffusion-enhanced and gravitational coagulation are also significant, as they increase the Brownian coagulation rates by 90% and 20%, respectively. Combined together, the coagulation processes can produce a significant decrease in mean column depth opacity of up to 40%, if $r_{eff,lift}$=1 μm, and 15%, if $r_{eff,lift}$=3 μm.

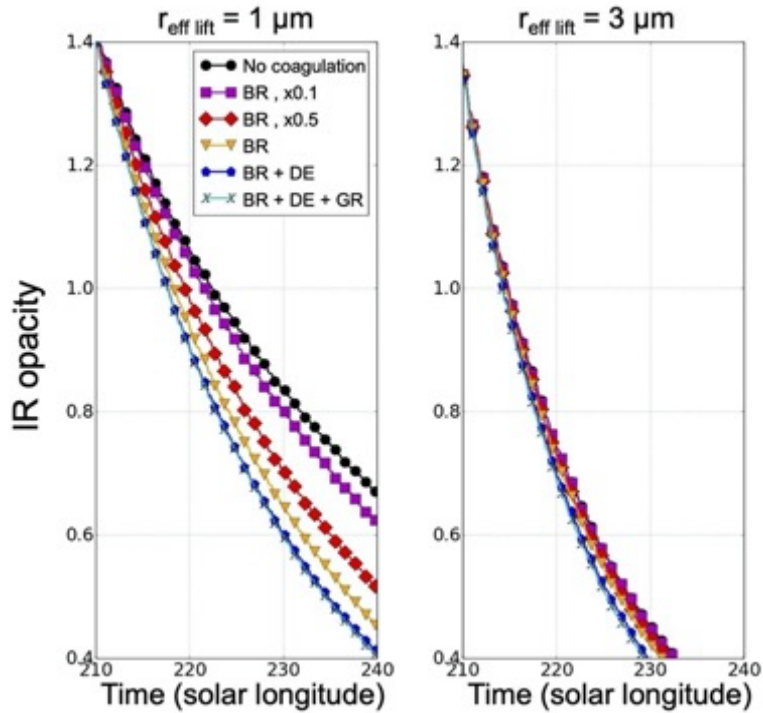

*Figure 15: The simulated storm's decay of IR opacity for different scenarios: reference case without active coagulation (black circles), with active Brownian coagulation only and coalescence efficiency $E_{coal}$=0.1 (purple squares), $E_{coal}$=0.5 (red diamonds), $E_{coal}$=1 (orange triangles), with both Brownian and Brownian diffusion-enhanced (blue pentagons), and with Brownian, Brownian diffusion-enhanced and gravitational coagulation (cyan crosses). The simulations are performed with a lifted particle size $r_{eff,lift}$ = 1 μm (left) and $r_{eff,lift}$ = 3 μm (right).*

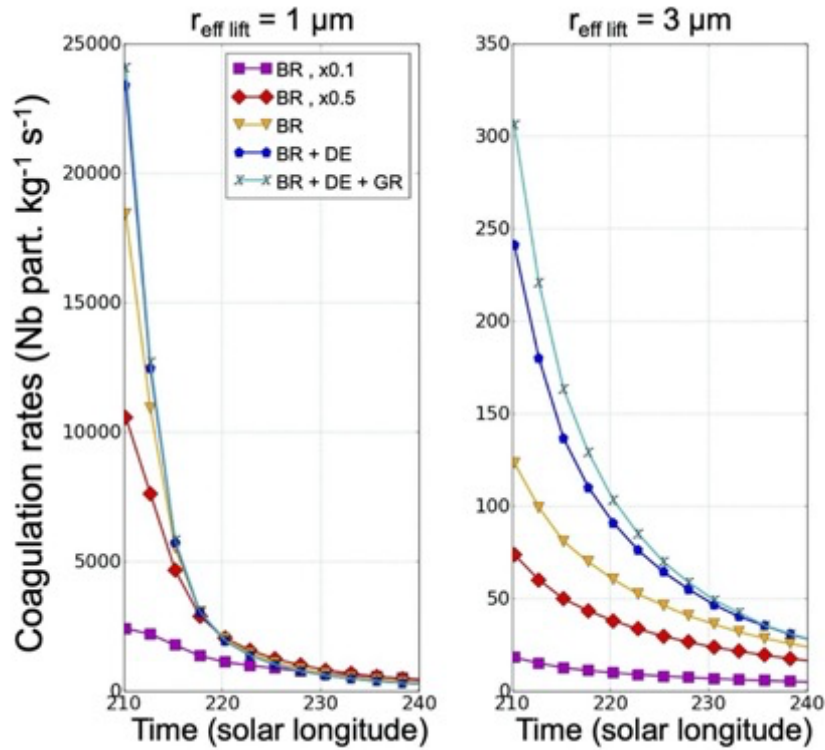

*Figure 16: as Figure 15, but showing the coagulation rates at 20 km altitude (loss of particle number/kg/s).*

# 5. Discussions

Coagulation of dust in the Martian atmosphere has not been directly observed, and therefore it is difficult to know whether this process actually occurs in the atmosphere. By using the MGCM coupled to a parameterization for dust coagulation and reasonable assumptions, we have been able to simulate Mars' climate with active dust coagulation while remaining consistent with available observations. We have shown that coagulation would occur on Mars and would have an appreciable effect during global dust storms. However, some caution is warranted, as many uncertainties remain regarding the mechanisms involved. Some of these approximations are considered in the following discussion.

First, our reference simulations did not include the full parameterization of external effects (van der Waals, viscous, Coulomb forces), of the fractal case, and of the correction factor $\omega_s$ (because of the large uncertainties associated with these effects on Mars). Instead, our simulations assumed that their net impact on dust coagulation balances each other (see Section 3.3.5), and were therefore performed with a coalescence factor $E_{coal} = 1$, which may lead to an overestimation of the coagulation efficiency. We tested a low-sticking scenario with a coalescence efficiency $E_{coal}$ set to 0.1, which leads to 10x lower coagulation rates and slower decay in opacity after the storm's peak, as the magnitude of the atmospheric response to coagulation scales with the efficiency. Our reference simulations were also performed with a relatively high collection efficiency $E_{g\,i,j}$ for gravitational coagulation, but the overall effect of this parameter is relatively small.

Second, our model does not take into account the change in dust particle shape that would be associated with coagulation. This would impact the sedimentation rates and the radiative transfer (due to changes in optical properties, in particular the forward scattering cross section would get smaller for fractal particles).

Third, our simulations did not include water ice clouds. Subgrid-scale processes such as "rocket dust storms" and slope winds (e.g. Wang et al., 2018), able to inject dust at high altitude and form detached layers of dust as observed, are not taken into account either. Exploring the impact of dust coagulation while taking into account these processes will be the topic of a further work.

Nevertheless, our results highlight the effects dust coagulation can have during major dust events on the atmosphere, storm evolution and particle sizes. In particular, we show that the coagulation rates are strongly sensitive to the number of small sub-micron particles, and therefore could be significantly increased if larger numbers of small particles are initially present in the atmosphere than what we considered in this study, for instance due to larger numbers of small particles lifted from the surface. Therefore, the impact of dust coagulation on Mars' climate may be larger than that presented in this paper.

It is remarkable to see that, when coagulation is active, the decay in opacity during the storm is not strongly sensitive to the effective radius $r_{eff,lift}$ of the lifted lognormal dust particle size distribution, as coagulation would allow for micron-sized dust particles to exist in the atmosphere even if only submicron-sized particles are lifted ($r_{eff}$ was increased from 0.7 to 1.4 µm in our simulation case with $r_{eff,lift}$=1 µm; Figure 6). Coagulation therefore allows us to lift large amounts of small particles while remaining consistent with observed column opacities.

GCMs simulating Mars' atmosphere typically use $r_{eff,lift}$=2-3 µm to inject dust from the surface into the atmosphere, so that the simulated atmospheric effective radii are close to those observed, i.e. 1-1.5 µm (and so that the subsequent simulated temperatures and storm's decay in opacity are consistent with observations). However, we have shown in this paper that a simulation with smaller injected particles ($r_{eff,lift}$=1 µm) and with active coagulation could also lead to atmospheric effective radii close to 1 µm, in particular during dust storms but also during quieter seasons, as shown by Figure 14.

Consequently, GCM studies could consider dust effective radii smaller than 2 µm for the lifting from the surface with active coagulation, and obtain results that would remain consistent with observations and even improve the agreement. In fact, many observations suggest the presence of a bimodal size distribution for dust (involving a population of submicron-sized particles, see Section 2). Injecting smaller particles would also lead to dust transport to higher altitudes, which may agree better with observations. Finally, key interactions with water ice clouds could also occur if a larger number of submicron-sized particles are present in the atmosphere. This cannot be explored in detail without active coagulation, otherwise small dust particles would accumulate in the atmosphere and produce unrealistically slow decay of dust following dust storms.

# 6. Summary and conclusions

We implemented the equations of coagulation (Brownian motion, Brownian diffusion enhancement, gravitational collection; Jacobson, 2005) for dust in the Ames Mars Global Climate Model, and explored the effects of this process during the 2018 global dust storm.

Our main results are:

- During dust storms, small dust particles could accrete to large particles by coagulation, at a rapid enough rate to offset their preferred sedimentation. This would lead to a global increase in the effective particle size in the atmosphere during the storm development and maintain a constant effective particle size during the decay phase. This was also shown by previous studies (e.g. Rossow (1978), Murphy et al. (1990), see Section 3.2).

- The effects of coagulation strongly depend on the number of small (sub-micron) particles and therefore on the choice of the lifted particle size distribution $r_{eff,lift}$. The process is especially critical when the effective particle size of dust lifted from the surface is defined to be small (e.g., 1 um). Typically, during a global storm, the particle effective radius in the atmosphere can be increased by a factor of 2, from 0.7 µm to 1.4 µm in the case $r_{eff,lift}=1$ µm (which involves large numbers of sub-micron particles), and by a factor of 1.3, from 1.8 µm to 2.4 µm in the case $r_{eff,lift}=3$ µm, thus leading to a faster decay after the storm's peak.

- Simulations performed with $r_{eff,lift}=1$, 2 and 3 µm and with coagulation produce a remarkably similar decay phase in IR opacity, which is not the case when coagulation is not considered (e.g., excess dust buildup would be obtained with $r_{eff,lift}=1$ µm due to low sedimentation rates). This is because during the storm, the small sub-micron particles are quickly removed by coagulation, and particle effective radii in the atmosphere quickly reach values greater than 1 µm. Consequently, the decay rate of the global storm cannot provide us with a strong theoretical constraint on the dust particle size in the atmosphere if we assume that coagulation occurs in the Martian atmosphere. In general, the representation of the decay phase of the storm relative to MCS dust observations is improved.

- The coagulation rates are maximum at Ls=210° (storm peak) and remain significant during the period Ls=190°-240°. IR 9.3 µm column dust opacities during this period are larger than 0.6, which could be seen as a threshold above which coagulation of dust must be taken into account in GCMs (this would include regional and global storms only).

- Coagulation rates are maximum at low altitude (higher pressure and larger amounts of dust) and in the tropical regions where dust is confined at this season.

- Coagulation due to Brownian motion with Brownian diffusion enhancement dominates coagulation due to gravitational collection. Gravitational coagulation may increase the coagulation rates (by up to 20%) during the storm's peak but only if large particles are involved ($r_{eff}=2$ µm in the atmosphere, which requires a lifted particle size $r_{eff, lift}=3$ µm).

- The effect of coagulation on the dust distribution, winds and temperature depends on the initial amount of small sub-micron particles present in the atmosphere (significant if $r_{eff,lift}=1$ µm, limited if $r_{eff,lift}=3$ µm). Overall, the upper atmosphere tends to be more depleted in dust (more easily removed through sedimentation as coagulation acts to increase the effective particle size) and colder (up to 20 K if $r_{eff,lift}=1$ µm, and 5 K if $r_{eff,lift}=3$ µm) during the mature and decay phases of the storm. The effects are negligible outside the storm period if $r_{eff,lift}=3$ µm, but are still significant if $r_{eff,lift}=1$ µm, as coagulation keeps the effective radius in the atmosphere close to 1 µm (instead of 0.5 µm without coagulation).

- GCM studies should consider lifting larger amounts of small dust particles from the surface. With active coagulation, the resulting atmospheric particle size would not strongly diverge from observations (as it would be the case without coagulation), while allowing dust to be transported at higher altitudes.

Our work opens the door to further studies of dust coagulation in climate models of Mars. First, coagulation remains to be explored with a broader spectrum of injected dust particle sizes in the model. For instance, we plan to study the impact of bi-modal dust distribution (involving a large micron-sized mode and a small submicron-sized mode) on Mars' climate with active coagulation, and better constrain particle sizes on Mars by comparing modeling results with observations. Second, in the MGCM, dust is slightly more confined to the low altitudes than what the observations suggest. In particular, observed detached layers of dust are not well reproduced. Coagulation remains therefore to be explored with more realistic vertical distributions for dust as well. Overall, further comparisons with observations are needed to better understand the impact of small particles and coagulation on Mars' climate. Finally, coagulation would also be an important process to consider when simulating past climates of Mars at high obliquity, during which the atmosphere has been shown to be thicker and able to maintain larger amounts of airborne dust (e.g., Newman et al., 2005). Mars-like exoplanetary atmospheres should also take into account this process if large amounts of dust are involved.

# Acknowledgements

T. B. was supported for this research by an appointment to the National Aeronautics and Space Administration (NASA) Postdoctoral Program at the Ames Research Center administered by Universities Space Research Association (USRA) through a contract with NASA.

# Appendix A - List of variables

- V : Dust particle volume (m$^{-3}$)
- V$_{rat}$ : Constant volume ratio
- r : Dust particle radius (m)
- g : Gravity constant (m s$^{-2}$)
- η : Atmospheric dynamic viscosity (Pa s or kg m$^{-1}$ s$^{-1}$), computed following the Sutherland's formula:

$$\eta(T) = \eta_0 \left(\frac{T}{T_0}\right)^{\frac{3}{2}} \frac{(T_0 + S)}{(T + S)}$$

  Where $T_0 = 273$ K, and $\eta_0 = 1.37 \times 10^{-5}$ kg m$^{-1}$ s$^{-1}$, S = 222 K in Mars conditions and $\eta_0 = 1.716 \times 10^{-5}$ kg m$^{-1}$ s$^{-1}$, S=111 K in terrestrial conditions.
- k$_b$ : Boltzmann constant ($\approx 1.38 \times 10^{-23}$ J K$^{-1}$)
- ε$_d$ : rate of dissipation of turbulent kinetic energy per gram of medium (m$^2$ s$^{-3}$),
- n : Dust particle number density (particle m$^{-3}$)
- T : temperature (K)
- ρ$_a$ : Atmospheric density (kg m$^{-3}$)
- ρ$_p$ : Dust particle density (kg m$^{-3}$)
- m : Molecular mass of a given gas (m$_a$) or particle (m$_p$) (kg)
- N$_B$ : Total number of size bins
- K$_{i,j}$ : Collision rate coefficient of two colliding particles of size bin i and j (m$^3$ part$^{-1}$ s$^{-1}$)
- E$_{coal}$ : Sticking (or coalescence) efficiency coefficient
- β$_{i,j}$ : coagulation rate coefficient of two colliding particles of size bin i and j (m$^3$ part$^{-1}$ s$^{-1}$)

$$\beta_{i,j} = E_{coal} K_{i,j}$$

- γ : Average thermal speed of an air molecule (γ$_a$) or particle (γ$_p$) (m s$^{-1}$)

$$\gamma = \sqrt{\frac{8 k_b T}{\pi m}}$$

- ν : Kinematic viscosity of air (m$^2$ s$^{-1}$)
- λ : Mean free path of a particle (m), given by λ=2ν/γ
- Kn : Knudsen number (dimensionless) of a particle or radius r, defined by: Kn=λ/r
- V$_{f,i}$ : Terminal fall speed of particle of size bin i (m s$^{-1}$), given by the Cunningham Stokes relationship (Pruppacher and Klett, 1997):

$$V_{f,i} = \frac{2}{9} \frac{r_i^2 \rho_p g}{\eta} G_i$$

  Where G$_i$ is the "slip flow" correction factor:

$$G_i = 1 + Kn_i \left[1.246 + 0.42 \exp\left(\frac{-0.87}{Kn_i}\right)\right]$$

- Re : Particle Reynolds number (dimensionless), given by Re=2r$_i$V$_{f,i}$/ν
- f$_{i,j,k}$: Volume fraction of V$_i$ + V$_j$ that is partitioned to each model bin k, given by:

$$f_{i,j,k} = \begin{cases} \left(\dfrac{V_{k+1} - (V_i + V_j)}{V_{k+1} - V_k}\right)\dfrac{V_k}{V_i + V_j} & V_k \leq V_i + V_j < V_{k+1} \quad k < N_B \\ 1 - f_{i,j,k-1} & V_{k-1} \leq V_i + V_j < V_k \quad k > 1 \\ 1 & V_i + V_j > V_k \quad k = N_B \\ 0 & \text{all other cases} \end{cases}$$

- $D_{p,i}$ : Brownian diffusion coefficient of particle of size bin i (m$^2$ s$^{-1}$)

$$D_{p,i} = \frac{k_b T}{6\pi r_i \eta} G_i$$

- $\delta_i$ : Mean distance (m) related to the mean free path of particle of size bin i, given by:

$$\delta_i = \frac{(2r_i + \psi_{p,i})^3 - (4r_i^2 + \psi_{p,i}^2)^{3/2}}{6 r_i \psi_{p,i}} - 2r_i$$

$$\psi_{p,i} = \frac{8 D_{p,i}}{\pi \gamma_p}$$

- $Sc_{p,i}$ : Particle Schmidt number, given by:

$$Sc_{p,i} = \frac{\nu}{D_{p,i}}$$

# Appendix B - Particle flow regime

The collisions, and therefore coagulations rates of particles, depend on both the atmospheric state and the particle flow regime (via the coagulation kernel parameter β in Equation 2), which can be described by the Knudsen number of air (which depends on the mean free path and thermal speed of an air molecule), and the particle Reynolds number.

|  | Symbol | Earth | Mars | Unit |
|---|---|---|---|---|
| Gravity | $g$ | 9.81 | 3.71 | m s$^{-2}$ |
| Dust particle density | $\rho_p$ | 2500 | 2500 | kg m$^{-3}$ |
| **Atmospheric state** | | | | |
| Dynamic viscosity at 273.15 K | $\eta_0$ | 1.72x10$^{-5}$ | 1.37x10$^{-5}$ | Pa s |
| Molar mass | $M$ | 28.97x10$^{-3}$ | 44.01x10$^{-3}$ | kg mol$^{-1}$ |
| Air temperature T | $T$ | 298 | 220 | K |
| Atmospheric density at 20 km | $\rho_a$ | 0.2 | 0.002 | kg m$^{-3}$ |
| **At air temperature T and density $\rho_a$** | | | | |
| Dynamic viscosity | $\eta$ | 1.84x10$^{-5}$ | 1.11x10$^{-5}$ | Pa s |
| kinematic viscosity | $\nu$ | 9.18x10$^{-5}$ | 5.55x10$^{-3}$ | m$^2$ s$^{-1}$ |
| Mean free path | $\lambda$ | 3.94x10$^{-7}$ | 3.41x10$^{-5}$ | m |
| Thermal velocity | $\gamma$ | 467 | 325 | m s$^{-1}$ |
| **For a 1 μm dust particle** | | | | |
| Knudsen number | Kn | 0.39 | 34.01 | |
| Reynold number | Re | 9.74x10$^{-6}$ | 3.85x10$^{-6}$ | |
| Particle mean free path | $\lambda_p$ | 1.14x10$^{-8}$ | 6.18x10$^{-7}$ | m |
| Particle mean distance | $\delta$ | 5.73x10$^{-9}$ | 3.65x10$^{-7}$ | m |
| Particle fall velocity | $V_f$ | 4.47x10$^{-4}$ | 1.07x10$^{-2}$ | m s$^{-1}$ |

*Table B-1: Atmospheric state and dust particle parameters used in this paper to compare the effect of dust coagulation in Earth's and Mars' atmospheres.*

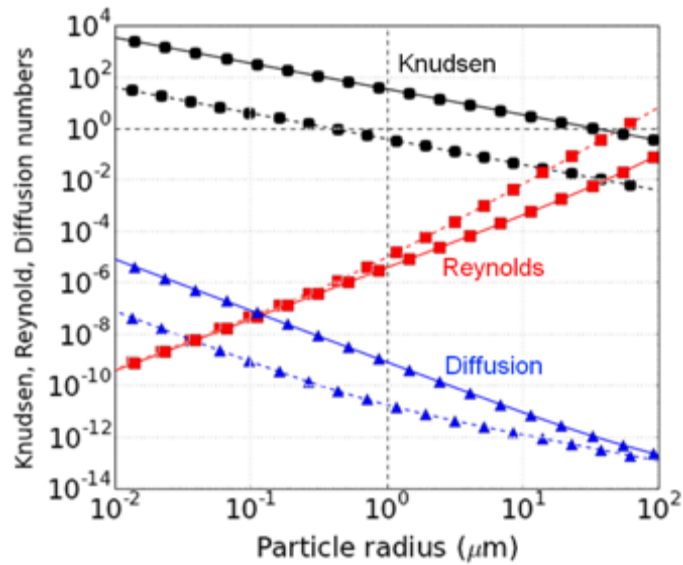

*Figure B-1: Knudsen number (black circles), Reynolds number (red squares) and particle diffusion coefficient (m$^2$ s$^{-1}$; blue triangles) for different sizes of falling dust particles (of density $\rho_p$ = 2500 kg m$^{-3}$) in Mars' (solid lines) and Earth's (dashed lines) atmospheres, with the following conditions: atmospheric temperature T = 220 K and air density $\rho_a$ = 0.002 kg m$^{-3}$ for Mars (i.e. typically ~20 km altitude during Fall equinox), T = 298 K and $\rho_a$ = 0.2 kg m$^{-3}$ for Earth (~20 km altitude). Small falling particles have a low terminal fall velocity and therefore a low Reynolds number.*

The different parameters used in this section are summarized for the Earth and Mars cases in Table B-1 and detailed in Appendix A. With Martian atmospheric conditions of T=220 K and air density $\rho_a = 0.002$ kg m$^{-3}$, we obtain an atmospheric kinematic viscosity $\nu=5.55 \times 10^{-3}$ m$^2$ s$^{-1}$, a thermal average thermal speed of an air molecule (CO$_2$) $\gamma=325$ m s$^{-1}$ and a mean free path $\lambda=3.41 \times 10^{-5}$ m. The (dimensionless) Knudsen number Kn$_i$ represents the ratio between the mean free path of an air molecule and the particle of size i. Kn<<1 (i.e. a large particle relative to the mean free path of air molecules) corresponds to the continuum regime, where the particle resistance to motion is due primarily to air viscosity. Kn>>10 (i.e. the mean free path of air molecules is large relative to the particle size) corresponds to the free molecular regime, in which the particle resistance to motion is due primarily to the inertia of air molecules hitting it. As shown in Figure B-1, the Knudsen number Kn on Mars varies from ~500 for a 0.1 µm dust particle to ~5 for a 10 µm dust particle (Kn > 50 for particles smaller than 1 µm). Therefore the particle flow regime is mostly in the free-molecular and transition regime (between the continuum and the free molecular regime). This remains true at all altitudes of the Martian atmosphere. Compared to the Earth's atmosphere, Mars' atmosphere has lower temperatures and thus air molecules with lower kinetic energies and average speeds. The roughly x100 lower atmospheric density also leads to a higher kinematic viscosity and a x100 longer mean free path for the air molecules. This produces a higher Knudsen number for dust particles on Mars than for the same particles on Earth, by a factor of ~100 (related to the ~100x lower atmospheric density on Mars in this case).

The particle Reynolds (dimensionless) number Re is the ratio of the inertial force exerted by the particle to the viscous force exerted by the air. It depends on 1) the terminal fall speed of the particle, 2) the particle radius, and 3) the "slip flow" correction factor G (see Appendix A). This correction factor is needed in the equation to take into account the fact that dust particles on Mars slip through the air with little viscous resistance. The Reynolds number of large dust particles (> 10 µm) is higher in terrestrial conditions (by a factor of ~100) due to the difference in atmospheric density (the fall velocity of such large particles is controlled by the difference in gravity and thus remains of the same order of magnitude on Earth and on Mars). On the other hand, small dust particles have a higher Knudsen number, a higher slip-flow correction term, and higher fall velocities in Martian than in terrestrial corrections, which balances the difference in density. As a result, one finds relatively similar values of Re, as shown on Figure B-1 for small particles.

# Appendix C - External forces impacting coagulation

Here, we intend to quantitatively estimate the impact of Van der Waals, viscous and Coulomb (electrical) forces on the coagulation of dust particles. Note that these forces can enhance or decrease the coagulation rate and can be treated either in terms of a collision kernel K or in terms of a coalescence efficiency $E_{coal}$.

## C.1. Van der Waals and viscous forces

Van der Waals forces are at the origin of contact forces between solids and result from fluctuations in the electron cloud of uncharged, nonpolar molecules, leading to the formation of momentary dipoles which can attract similar dipoles in other molecules. In our case, the Van der Waals' interaction energy between spherical dust particles was approximated by Hamaker (1937) and given by equation 13.A.9 in Seinfeld and Pandis (2006). It depends on the Hamaker constant $A_H$ that depends on the material properties (it can be positive or negative in sign depending on the intervening medium). The correction factor for coagulation due to the van der Waals forces is given by equation 13.A.14 in Seinfeld and Pandis (2006). By using this equation with $A_H/kT$ ranging from 20 to 200, as suggested by Seinfeld and Pandis (2006), we reproduced their results obtained for terrestrial conditions (Seinfeld and Pandis, 2006; their figures 13.A.3 and 13.A.4). As shown by these authors, van der Waals forces enhance the rate of coagulation of small particles in the free-molecular regime. We applied the equations in the case of dust particles colliding in Mars' atmosphere and obtained similar results (Figure C-1), with a correction (enhancement) factor to Brownian coagulation typically ranging between 1.2 and 1.6. The enhancement is particularly strong (with a correction factor higher than 1.6) when particles smaller than 0.1 µm collide (not shown).

On the other hand, the resistance of the atmosphere as two particles approach each other induces viscous forces (i.e. fluid dynamic interactions between the particles and the atmosphere) which tend to retard the collision (and thus the coagulation) rate. The correction factor for coagulation due to the viscous forces is given by equations 13.A.19 and 13.A.20 in Seinfeld and Pandis (2006). We reproduced their results (their figure 13.A.4) for terrestrial conditions and found similar results for the case of dust particles in Martian conditions. As shown by Figure C-1, the combined effect of van der Waals + viscous forces could impact the coagulation rate by a factor ranging from 0.75 to 1.2, depending on $A_H$.

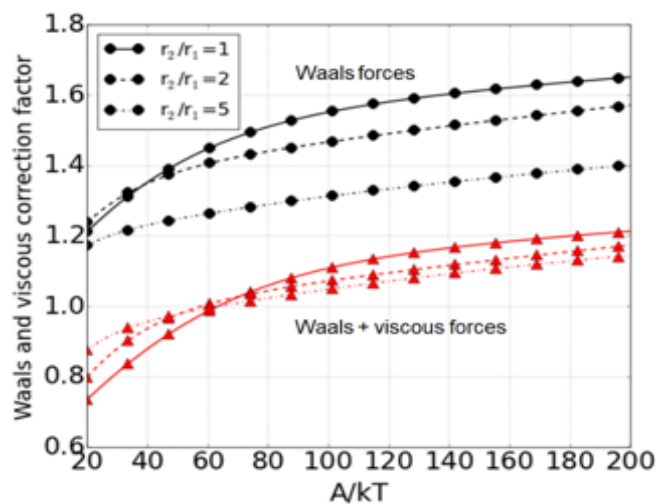

Figure C-1: *Correction factor (vs $A_H/kT$) to the Brownian coagulation coefficient when van der Waals only (black lines, disks) and van der Waals + viscous forces (red lines, triangles) are included in the calculation for the case of two dust particles colliding in Mars' atmosphere. Results are obtained for a ratio of particle radius of 1 (solid line), 2 (dashed) and 5 (dot-dashed).*

## C.2. Coulomb forces

Charged particles and their force field (so-called Coulomb forces) can also modify the collision frequency and thus alter the sticking efficiency between particles. Charged particles could experience either enhanced or retarded coagulation rates depending on their charges. The correction factor for coagulation due to the Coulomb forces is given by equation 13.A.16 in Seinfeld and Pandis (2006) and depends on the particle charges and the dielectric constant of the medium. In the case of dust particles in Mars' atmosphere, the absence of constraints on these parameters and the absence of detailed modeling makes it difficult to estimate the effect of Coulomb forces on coagulation. Although we can expect two particles of similar size to carry the same amount of charges and thus to repel each other (which would reduce their accretion efficiency), the net result is uncertain for two particles of different sizes. However, as in the terrestrial case (Seinfeld and Pandis, 2006), the correction factor for Coulomb forces is expected to remain within ±50%.

# Appendix D - Coagulation of fractal dust particles

The collision kernels presented in Section 3.3.3 assumed spherical dust particles. However, some dust particles could be fractal (e.g. chain-like aggregates), which impacts the collision and coagulation rates. The effect of fractal geometry is treated for Brownian coagulation by Jacobson (2005; see their equation 15.52) by replacing the volume-equivalent radius $r_i$ in the collision kernel $K^B$ (Equation 4) by a mobility radius in the diffusion $D_{p,i}$, mean distance $\delta_i$ and thermal velocity $\varphi_{p,j}$ terms, and by the fractal radius $r_f$ of the agglomerate elsewhere:

$$r_{f,i} = r_s N_i^{1/D_f} \qquad (D1)$$

Where $r_s$ is the radius of each individual monomer (assumed to have the same size), $D_f$ the fractal dimension ($D_f$=3 for pure sphere), and N the total number of monomers in the particle.

Models suggest that dust particles on Mars could have a characteristic fractal dimension ranging from 2 to 3 with a few monomers (Ding et al., 2020, Hou et al., 2018). Figure D-1 shows how the coagulation rate changes when fractal dust particles are assumed. Assuming a lower fractal dimension $D_f$ enhances the coagulation rate of dust particles, except maybe for large particles (> 2 µm), while assuming a lower monomer size $r_s$ (i.e. a larger number of monomers) enhances the coagulation rate in all cases.

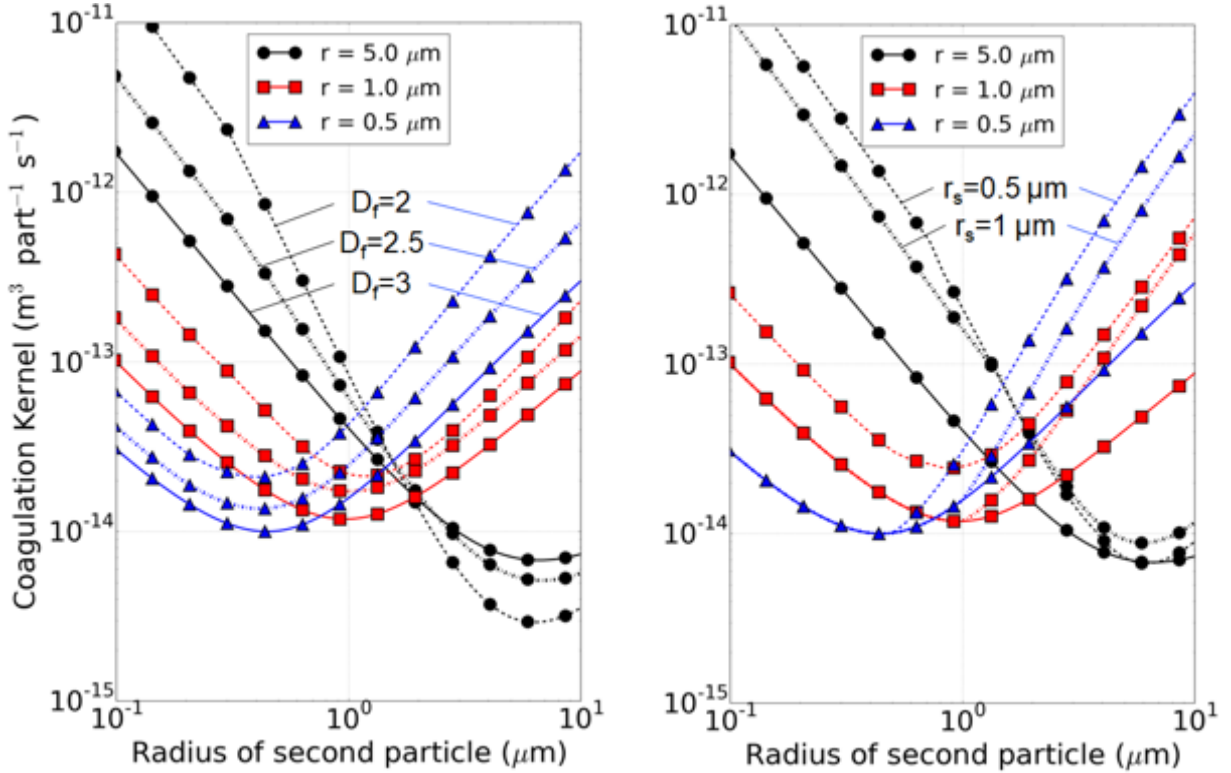

*Figure D-1: Brownian coagulation kernel ($m^3$ particle$^{-1}$ s$^{-1}$) for 3 dust particles of volume-sphere equivalent radius 5 µm (black disks), 1 µm (red squares), 0.5 µm (blue triangles) colliding with other dust particles in the range 0.1-10 µm (x-axis) in Mars' atmosphere. Left: Reference spherical case (as in Figure 1, solid line, equivalent to fractal dimension $D_f$=3) compared with fractal cases with monomer size $r_s$=0.2 µm and $D_f$=2.5 (dot-dashed) and $D_f$=2 (dashed). Right: Reference spherical case compared with fractal cases with $D_f$=2 and monomer size $r_s$=0.5 µm (dot-dashed) and 1 µm (dashed).*

# Appendix E - Numerical coagulation tests in 0-D on lognormal dust particle size distributions in Martian conditions

Here we estimate the stability of a lognormal dust particle size distribution to coagulation only, in 0-dimension (we assume one finite atmospheric volume with no source and no sink for the total mass of dust particles), in Martian conditions, as in Fedorova et al., 2014 (see their section 6.4 and Fig. 16). This enables us to test the response of a lognormal particle size distribution to coagulation (and focus on characterizing the coagulation process only). Note that we were able to reproduce the results of Fedorova et al. (2014), which is expected.

We consider the Brownian, Brownian diffusion-enhanced and gravitational coagulation and solve the equations presented in Section 3 and based on the semi-implicit scheme described in Jacobson (2005), with conservation of mass and volume (equation 3). The parameters of this 0-D coagulation model are given by Table B-1 (Mars atmospheric state) and Table E-1 (model settings). We test the algorithm presented in Section 4.1.2, in which the particle size distribution is forced to remain lognormal at every timestep, with a fixed variance, as an approximation for convenience in coding in the MGCM. Finally, we evaluate the error made with this approximation on the effective radius and effective variance of the particle size distribution.

| Parameters for the 0-D simulations | |
|---|---|
| Minimum radius of the size distribution | $r_{min}$ = 0.005 μm |
| Maximum radius of the size distribution | $r_{max}$ = 20 μm |
| Number of size bins | $N_B$ = 40 |
| Simulation length | L = 100 days |
| Timestep | h = 0.1 sol |
| Initial effective radius | $r_{eff}$ = 1 μm |
| Initial effective variance | $\sigma_{eff}$ = 0.5 μm |
| Initial particle number density | $N_{0\ non-storm}$ = 6 cm$^{-3}$<br>$N_{0\ storm}$ = 100 cm$^{-3}$ |

*Table E-1: Settings and parameters for the 0-D coagulation model. The other model settings and atmospheric state are given by Table B-1 (Mars conditions).*

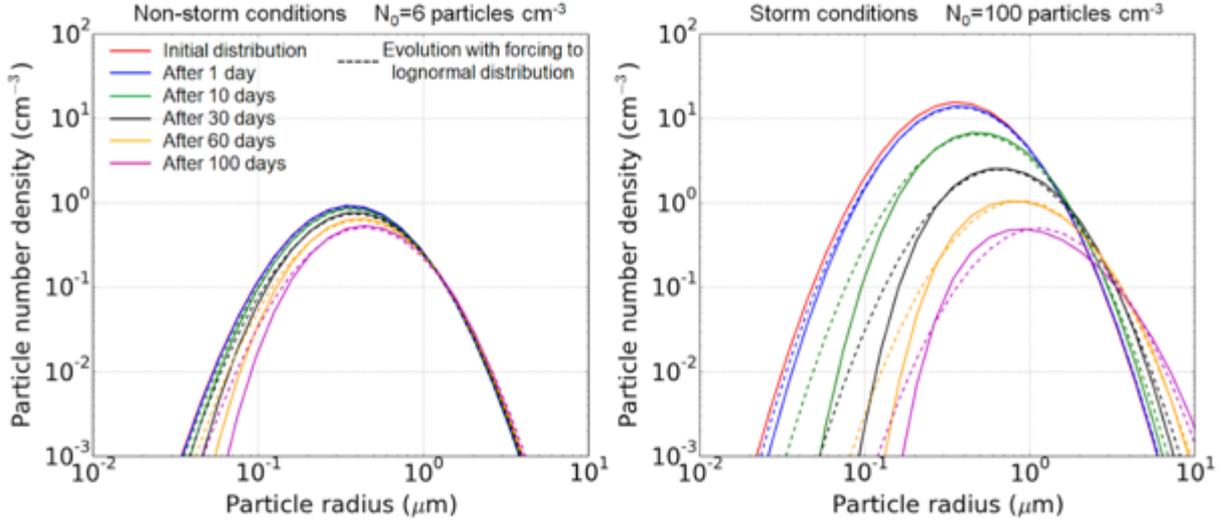

*Figure E-1: Evolution of an initially lognormal dust particle size distribution with coagulation during 100 days (solid lines), at ~20 km altitude in the Martian atmosphere (Fall equinox, T=220 K, $\rho_a$=0.002 kg m$^{-3}$). The atmospheric state and model settings are summarized in Table B-1 and Table E-1, respectively. The initial distribution (red) has $r_{eff}$=1 µm and $\sigma_{eff}$=0.5 (this corresponds to a mean radius $r_0$=0.36 µm and mean variance $\sigma_0$=0.64). The particle density is $\rho_d$=2500 kg m$^{-3}$. A: In non-storm condition, with dust particle density N=6 particles cm$^{-3}$. B: In storm condition, with N=100 particles cm$^{-3}$. The dashed lines show the evolution of the particle size distribution with coagulation but with the distribution being forced to remain lognormal at each timestep, with a fixed variance $\sigma_{eff}$ = 0.5, as detailed in Section 4.1.2. The different colors mark the number of days from the beginning, as indicated by the legend.*

Figure E-1 shows the evolution of the initially lognormal particle size distribution (described in Table E-1) on a timescale of 100 Earth days, in the Martian atmosphere at about 20 km altitude in non-storm (particle density N=6 particles cm$^{-3}$) and storm (N=100 particles cm$^{-3}$) conditions. Table E-2 summarizes the results of these 0-D calculations.

In non-storm conditions, the particle distribution is relatively stable during the first 10 days and starts to significantly decrease in concentration after this period (as the small particles accrete to the larger ones). After 30 and 100 days, the initial population peak decreased by 20% and 46%, respectively, while $r_{eff}$ increased from 1 µm to 1.06 µm and 1.18 µm, respectively. The variance decreased by only ~10% after 100 days.

In storm conditions, the initial particle number density ($N_0$=100 cm$^{-3}$, i.e. 10x more particles than in the non-storm case) falls much faster, as the rate of loss of small particles increases with the initial number density $N_0$ of the lognormal distribution. $N_0$ decreased by 12% after one day. After 10 days, it decreased by 58%, while particles with a radius r < 0.1 µm disappeared. After 30 and 100 days, the initial population peak decreased by 75% and 97% (only ~3.22 particles cm$^{-3}$ are left), respectively, while $r_{eff}$ increased from 1 µm to 1.76 µm and 3.38 µm, respectively. The variance remained about 0.5 by +/-10% over the 100 days.

| Time (Earth day) | $r_{eff\,mom}$ (µm) | $\sigma_{eff\,mom}$ | $N_{0\,mom}$ (cm$^{-3}$) | $r_{eff\,log}$ (µm) | $\sigma_{eff\,log}$ | $N_{0\,log}$ (cm$^{-3}$) | Error on $r_{eff}$ (%) | Error on $\sigma_{eff}$ (%) | Error on $N_0$ (%) |
|---|---|---|---|---|---|---|---|---|---|
| Non-storm conditions $N_0$ (t=0) = 6 cm$^{-3}$ ||||||||||
| 1 | 1.00 | 0.499 | 5.83 | 1.00 | 0.5 | 5.71 | 0.05 | 0.16 | -2.12 |
| 10 | 1.02 | 0.493 | 5.47 | 1.02 | 0.5 | 5.36 | 0.45 | 1.50 | -2.12 |
| 30 | 1.06 | 0.479 | 4.80 | 1.07 | 0.5 | 4.70 | 1.18 | 4.12 | -2.12 |
| 60 | 1.11 | 0.463 | 4.03 | 1.13 | 0.5 | 3.94 | 2.01 | 7.35 | -2.12 |
| 100 | 1.18 | 0.447 | 3.28 | 1.21 | 0.5 | 3.21 | 2.78 | 10.67 | -2.12 |
| Storm conditions $N_0$ (t=0) = 100 cm$^{-3}$ ||||||||||
| 1 | 1.03 | 0.489 | 87.7 | 1.04 | 0.5 | 85.9 | 0.58 | 2.25 | -2.12 |
| 10 | 1.28 | 0.431 | 42.0 | 1.33 | 0.5 | 41.1 | 3.13 | 13.73 | -2.12 |
| 30 | 1.76 | 0.406 | 16.1 | 1.82 | 0.5 | 15.8 | 3.36 | 18.79 | -2.12 |
| 60 | 2.43 | 0.442 | 6.83 | 2.43 | 0.5 | 6.69 | -0.21 | 11.68 | -2.12 |
| 100 | 3.38 | 0.543 | 3.22 | 3.12 | 0.5 | 3.15 | -8.46 | -8.54 | -2.12 |

*Table E-2: 0-D coagulation model results in Martian conditions, as shown by Figure E-1. Evolution over 100 Earth days of the effective radius, effective variance and particle number density of the initially lognormal particle size distribution with coagulation ($r_{eff\,mom}$, $\sigma_{eff\,mom}$, $N_{0\,mom}$), of the same parameters but with the approximation that the distribution remains lognormal at every timestep ($r_{eff\,log}$, $\sigma_{eff\,log}$, $N_{0\,log}$), with constant $\sigma_{eff\,log}$ = 0.5, and of the errors made with this approximation.*

The dashed line on Figure E-1 indicates the evolution of the particle size distribution when the distribution is forced to follow a lognormal function, with a fixed variance ($\sigma_{eff}$=0.5) at each timestep, as detailed in Section 4.1.2. This allows us to investigate the error made when simulating coagulation with the 3-D MGCM, which approximates all atmospheric particle size distributions with a lognormal distribution.

As shown by Figure E-1, the particle size distribution after a few days diverges from a lognormal distribution as the smallest particles are removed by coagulation. Table E-2 shows that after 30 days, the effective radius of the particle size distribution resulting from coagulation (calculated from its moments) is $r_{eff\,mom}$=1.06 µm and $r_{eff\,mom}$=1.76 µm in non-storm and storm conditions, respectively, while the fit to the lognormal gives $r_{eff\,log}$=1.07 µm and $r_{eff\,log}$=1.82 µm, respectively. The difference corresponds to an error of only ~1.2% and 3.3%, respectively. Similarly, the forcing to lognormal function produces an error on the particle number density that is ~2% over the entire simulation, compared to the case without forcing to lognormal distribution.

Larger errors of up to 19% can be realized for the variance after 30 days in storm conditions and of 8.5% on $r_{eff}$ after 100 days. Overall, these errors —and thus the lognormal approximation for a particle size distribution undergoing coagulation— remain acceptable for the purpose of this study, considering the fact they correspond to maximal errors under unlikely conditions (absence of continuing sources and sinks) and over long timescales. In the dynamic atmosphere of Mars, coagulation would be balanced by other processes, such as sedimentation (sink; removes the largest particles, preventing further collisions and aggregations with other particles), surface dust lifting or atmospheric vertical and horizontal dust transport (source of dust particles; brings new small particles, balancing the loss of small particles due to coagulation). These processes would occur on timescales that are much shorter than coagulation, and would play a big role in the maintenance of a lognormal distribution. Consequently, the effects of dust coagulation on Mars are better investigated with 3-D GCMs, which take into account the dynamics and physics of the Martian atmosphere. This investigation is performed in Section 4.2.